# Synthesis and Characterization of Two-dimensional $Cs_2AgInCl_6$ Nanoplates as Building Blocks for Functional Surfaces

*Sasha Khalfin[#1], Noam Veber[#1,2], Saar Shaek[1], Betty Shamaev[1,2], Shai Levy[1], Shaked Dror[1], Yaron Kauffmann[1], Maria Koifman Khristosov[1], and Yehonadav Bekenstein[1,2,3]*

1. Department of Materials Science and Engineering, Technion – Israel Institute of Technology, 3200003 Haifa, Israel.

2. The Solid-state institute, Technion – Israel Institute of Technology, 3200003 Haifa, Israel.

3. The Nancy and Stephen Grand Technion Energy Program, Technion – Israel Institute of Technology, 32000 Haifa, Israel.

*Corresponding author. Email: bekenstein@technion.ac.il

#These authors contributed equally

**Abstract:** breaking crystal symmetry is essential for engineering emissive double perovskite metal halides. The goal is to overcome their inherently indirect and disallowed optical transitions. Here we introduce a synthesis for silver - $Cs_2AgInCl_6$ two-dimensional hybrid nanoplate products that break the symmetry in two ways, their shape and their heterointerfaces. A comparative study between $Cs_2AgInCl_6$ nanocubes and nanoplates is presented to emphasize the difference in optical properties. A modified colloidal synthesis for $Cs_2AgInCl_6$ yields high-quality nanoplates with small lateral dimensions very different from the symmetric cubes. Each nanoplate is decorated with metallic silver nanoparticles, with diameters on the scale of the thickness of the perovskite nanoplate, forming significant heterointerfaces that further break symmetry. The $Cs_2AgInCl_6$ two-dimensional nanoplates also demonstrate facile transformation into larger crystalline nanosheets once deposited on substrates. We thus highlight those nanoplates as potential building blocks for assemblies of functional surfaces.

Lead-halide perovskite nanocrystals are well known for their bandgap tunability through anion exchange[1],[2], crystallographic transformations[3–5], and outstanding optoelectronic applications[6],[7,8]. However, the toxicity of lead-halide perovskites is of great concern in considering large-scale implementation. Replacing the lead in lead-halide perovskites with other elements while retaining the favorable optoelectronic properties is of practical interest and fundamental importance [9]. Understanding the properties of lead-free perovskites affects the resulting electronic structure and thermodynamic stability of the product, and is essential for the future development of these materials[10–13]. One alternative is a double-perovskite with the composition of $Cs_2AgInCl_6$, where $Pb^{2+}$ is replaced with alternating $Ag^{1+}$ and $In^{3+}$ in a doubled unit cell. The resulting product is highly crystalline, stable, and has a direct bandgap. Unfortunately, unlike the lead-halide compositions, it is not an efficient light emitter due to its symmetry-forbidden transition[12,14,15]. Methods to circumvent this problem, include metal-ion alloying or doping, which have demonstrated bright emission for doped $Cs_2AgInCl_6$ compositions[12,14,16–18]. Another striking difference between the lead-halide system and the double-perovskites is the excitonic confinement. Whereas lead-halide lower dimensionalities significantly affect optical properties, as shown in nanocubes[19], nanowires[20], nanoplates[21], and nanobelts[22]. In double-perovskites, minor electronic modifications are observed due to the localized self-trapped nature of the excitons, with a small Bohr radius. [14,15,23–26].

Here, we demonstrate $Cs_2AgInCl_6$ two-dimensional nanoplates structure, introducing an alternative symmetry-breaking method for modified electronic transitions. $Cs_2AgInCl_6$ two-dimensional nanoplates differ from their $Cs_2AgInCl_6$ nanocubes counterparts, both in morphology and by their silver nanoparticle decorations influence. These differences affect their electronic properties [27,28]. Each nanoplate is decorated with metallic silver nanoparticles, with diameters on the scale of the thickness of the perovskite nanoplate, resulting in a hybrid perovskite-silver structure.
The most studied semiconductor−metal hybrid systems contain CdS nanorods with metal nanoparticles, such as Ag or Au[29,30]. Lead-halide-Au and Lead-halide-AuCu hybrid nanoparticles were recently demonstrated. They show modified emission properties depending on the metallic island size and composition[31–35]. Since the coupling of halide perovskites and metallic particles is a relatively new field, it is fundamentally interesting to understand the extent of the modified properties of the hybrid structure. In this study, we present the results of our novel synthesis of $Cs_2AgInCl_6$ two-dimensional nanoplates and compare them to $Cs_2AgInCl_6$ double-perovskite nanocubes. The resulting products of the developed synthesis are $Cs_2AgInCl_6$ two-dimensional nanoplates with small lateral dimensions. This contrasts larger lateral dimension $Cs_2AgInCl_6$ nanoplates published recently by Ma et al.[36], which show no evidence of silver nanoparticles. In our study, which focuses on small nanoplates, the metallic part is more exposed to the surroundings compared to nanocubes, due to their minimal thickness. We hypothesize this open a door to possible applications that these nanoplates could have in the future, through the use of the properties of the metallic nanoparticles inherent in them. Furthermore, we observe that small nanoplates often evolve into extensive nanosheets when deposited on surfaces. This indicates a greater thermodynamic stability of the nanosheets on these surfaces, compared to the nanoplates. Such findings highlight the potential of these $Cs_2AgInCl_6$ two-dimensional nanoplates as foundational elements for intricate structural designs.

In this study, inspired by the synthesis of Bai et al.[37] for quadruple-perovskite nanocrystals, we developed a modified hot-injection method to synthesize a $Cs_2AgInCl_6$ two-dimensional nanoplates with small lateral dimensions. In our synthesis, the metal acetate (OAc) precursors (i.e., Cs(OAc), Ag(OAc), and $In(OAc)_3$), the organic ligands of oleic acid (OA), and oleylamine (OLA) were loaded into a 3-neck flask, along with ODE. This reaction mixture was heated to 110°C under a vacuum for 1 hour and changed its color from colorless to brown (which suggests the formation of silver nanoparticles). Then, the solution was heated to 170°C under a nitrogen atmosphere. Chlorotrimethylsilane (TMSCl) was swiftly injected when the temperature reached 170°C to initiate the nucleation and growth of $Cs_2AgInCl_6$ nanoplates. The reaction solution was continuously heated to 180°C and after 1 min at this temperature, immediately cooled using an ice bath to room temperature; see the schematics in Figure 1a and Figure S1. The solution was cleaned, and in the end, redispersed in hexane; see the insert of Figure 1a for a photo of product's typical solution. For additional details, see the Experimental Section in SI.

Essentially, for nanoplates synthesis, we used a less common Cl-precursor of trimethylsilyl chloride (TMS-Cl) in our synthesis (see SI for details). TMS-Cl is known as the initiative of faster nucleation and growth processes of nanocrystals, relative to Cs-oleate (which is common in nanocubes synthesis) [38–40]. The resulting product was a double-perovskite of $Cs_2AgInCl_6$ nanoplates composition, decorated by metallic silver nanoparticles. The $Cs_2AgInCl_6$ perovskite has a cubic unit cell with a space group of Fm3m, in which the $[AgCl_6]^{5-}$ and $[InCl_6]^{3-}$ octahedra form a 3D framework, see a polyhedral structural model in Figure 1b.

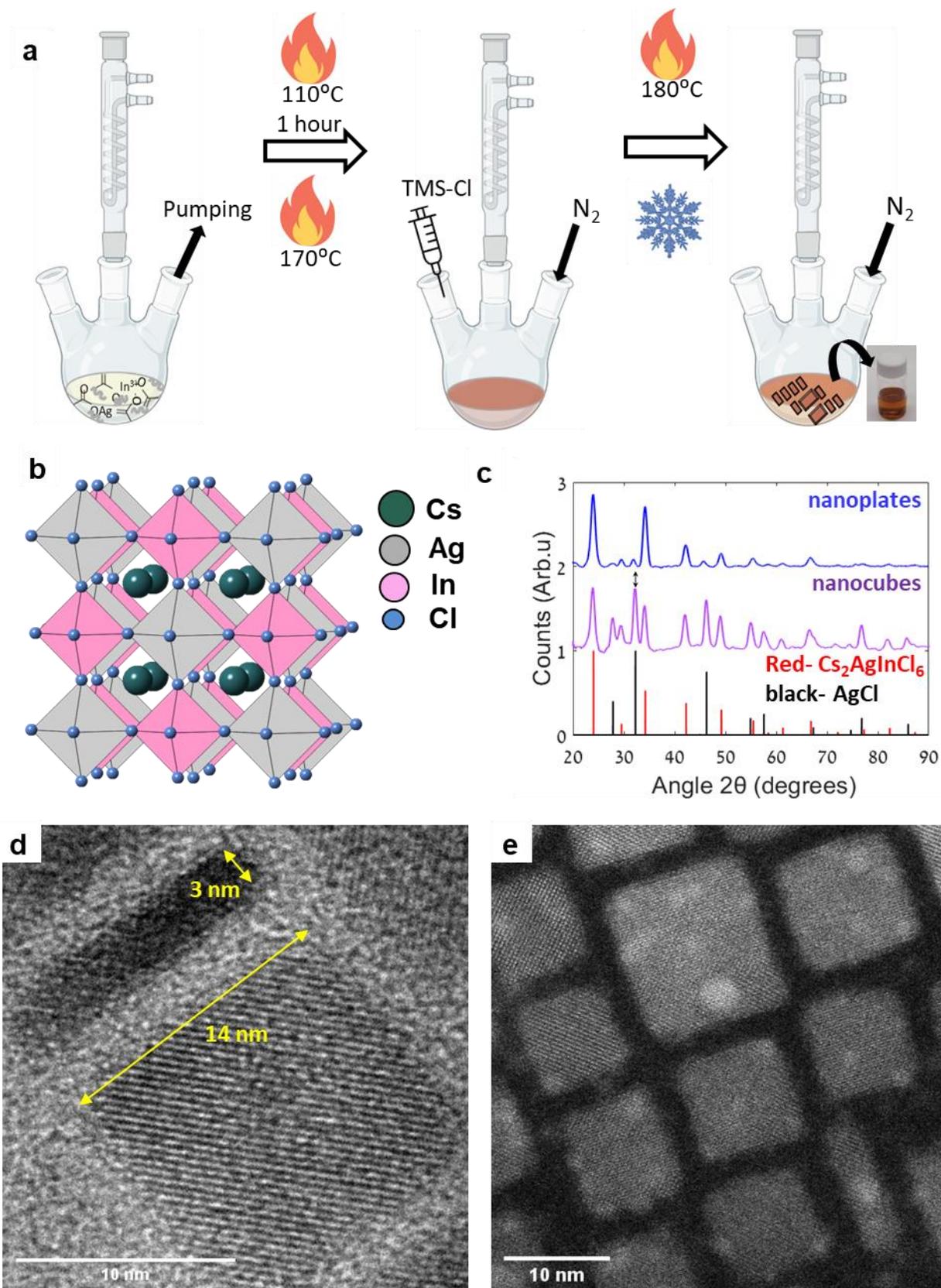

Figure 1: (a) Schematic synthesis procedure for the $Cs_2AgInCl_6$ double-perovskite nanoplates. Inset; a photo of a typical solution of $Cs_2AgInCl_6$ two-dimensional nanoplates.
(b) A polyhedral structural model of a double-perovskite with $Cs_2AgInCl_6$ composition.
(c) XRD structural characterization, displaying patterns of $Cs_2AgInCl_6$ nanoplates in comparison to

$Cs_2AgInCl_6$ nanocubes, with reference ICDD 01-085-7533 (red columns) and their AgCl by-product, with reference ICDD 00-001-1013 (black columns), measured with Cu source ($\lambda_0 = 1.5406$ Å). The two-way arrow pointing to the main peak of the AgCl in both nanocrystals emphasizes that in the synthesis of the nanoplates, there is significantly less AgCl by-product (by seven times). AgCl peaks widening emphasizes that the AgCl nanoparticles are formed during the synthesis.
(c) and (d) depicted a top view and a side view of $Cs_2AgInCl_6$ nanoplates, when (c) is a typical HR-TEM micrograph and (d) is a HAADF-STEM micrograph.

To verify the crystal structure, X-ray diffraction (XRD) study of the product was conducted and confirmed the presence of perovskite phase (ICDD no. 01-085-7533) as well as a silver chloride byproduct (ICDD no. 00-001-1013), as shown in Figure 1c.

Our XRD structural characterization displays patterns of $Cs_2AgInCl_6$ nanoplates compared to $Cs_2AgInCl_6$ nanocubes (see Extended Explanations -S3 in SI). The synthesis of $Cs_2AgInCl_6$ nanocubes is based on Dahl *et al.*, and Levy *et al.*, studies [24,41]; for details and Transmission Electron Microscope (TEM) characterization, see SI and Figure S2. This synthesis is simple to perform (no need of Schlenk-line) and obtains a typical product for the $Cs_2AgInCl_6$ nanocubes[24,25].

An immediate advantage of nanoplates synthesis over nanocubes synthesis [17,19] is reduced formation of AgCl by-products (by seven times), emphasizing the efficiency of our nanoplates developed synthesis. This is apparent from the comparison of AgCl main XRD peak in nanocubes and nanoplates (marked by an arrow in Figure 1c). The $Cs_2AgInCl_6$ nanoplates' lateral dimensions are 13.2±3.7nm, and the thickness is a few unit cells (~4.3±1.3 nm); see Figure S3(a-c) in SI. TEM micrographs depict $Cs_2AgInCl_6$ nanoplates in two orientations; A face down and a side up of stacked nanoplates (see Figure 1d-e and Figure 2a). The $Cs_2AgInCl_6$ nanoplates are decorated by high-contrast spherical metallic silver nanoparticles with lateral dimensions of 2.0±0.6 nm (see S3(d) in SI.), which are formed during the synthesis and don't change under TEM microscope observations. Selected area electron diffraction was taken from a typical sample area (as shown in Figure S3(a)) seen in Figure 2b, resulting in a polycrystalline ring pattern, indicating the $Cs_2AgInCl_6$ phase, as expected. To further probe the identity of higher contrast spherical decorations, we acquired high-angle annular dark-field scanning TEM (HAADF-STEM) micrographs, as shown in Figure 1e and Figure S4(a-b) in SI.

We also have evidence of beam-induced void formations in $Cs_2AgInCl_6$ nanoplates, as we reported in the past in $Cs_2AgInCl_6$ nanocubes [14,15,23–25], marked with a yellow arrow in Figure h(b), but we will not delve into that in this article. The energy dispersive X-ray spectroscopy (EDX) elemental mapping (see Figure S5 in SI) resulted in the following compositions of the perovskite itself: 19.5± 2.6% Cs, 9.1 ± 1.2% Ag, 8.6 ± 1.2% In, and 62.8 ±4.8% Cl (atomic fraction), resembling the expected 2:1:1:6 stoichiometry, in a strong agreement with previous reports for $Cs_2AgInCl_6$ [14,15,24].

In some cases, chlorine lack is detected (Figure 2c-f and in Figures S6). This phenomenon of halide lack in the characterization of perovskites has already been observed in the past[35]. It is caused by the low mass of chlorine and its weak connection to the edge of the octahedron. This phenomenon can be accelerated in nanoplates due to their large surface area compared to nanocubes. The EDX confirms that the spherical nanoparticles are metallic silver, see Figure 2d. This agrees with the EDX mapping by Dahl *et* al. and others [15,24].

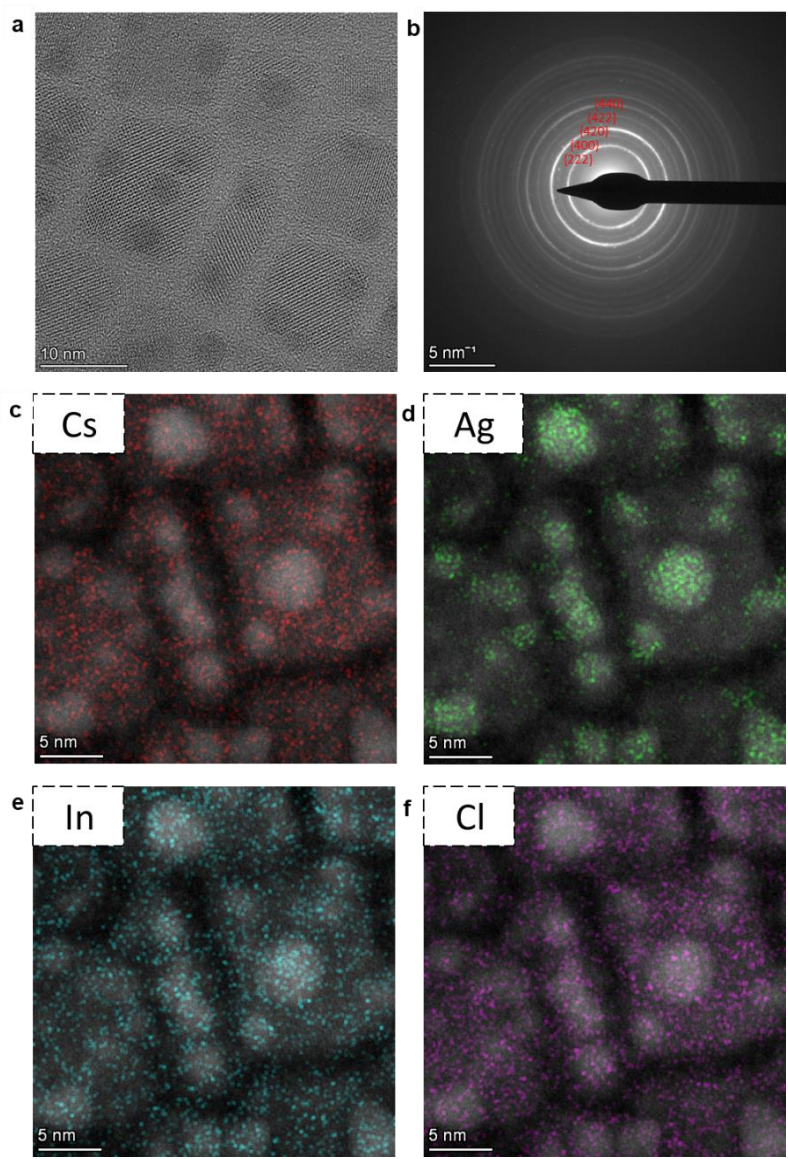

Figure 2: (a) Typical high-resolution TEM micrograph of $Cs_2AgInCl_6$ nanoplates with lateral dimensions of 13.2±3.7 nm and thickness of a few unit cells (~4.3±1.3 nm) with high-contrast spherical metallic silver nanoparticles decorations with lateral dimensions of 2.0±0.6.
(b) Selected area diffraction of $Cs_2AgInCl_6$ nanoplates presenting a polycrystalline ring pattern (ICDD 01-085-7533). (c)-(f) EDX elemental mapping of a typical area of $Cs_2AgInCl_6$ nanoplates for cesium, silver, indium, and chlorine, respectively. The silver nanoparticles are shown on a perovskite background.

To study $Cs_2AgInCl_6$ nanoplates' optical properties, their absorbance and photoluminescence (PL) spectra were measured. Figure 3(a) shows the absorbance spectra of $Cs_2AgInCl_6$ nanoplates dispersed in hexane (represented by blue lines) compared to $Cs_2AgInCl_6$ nanocubes dispersed in hexane (represented by purple lines). The absorbance spectrum of $Cs_2AgInCl_6$ nanocubes does not show any peak, as shown in our previous study [24]. However, $Cs_2AgInCl_6$ nanoplates peak metallic silver surface plasmon peak, which we saw before in $Cs_2AgInCl_6$ nanocubes only after an illumination experiment [24].

Metallic nanostructures, like silver nanoparticles, exhibit the phenomenon of localized surface plasmons, which are collective oscillations of the conduction electrons coupled to the electromagnetic field. The frequency and intensity of the oscillations are characteristic of the material type and are highly sensitive to interfaces with the surrounding medium and the nanostructure geometry[42–44]. The surface plasmon resonance of $Cs_2AgInCl_6$ two-dimensional nanoplates was observed previously on $Cs_2AgInCl_6$ nanocubes after photoinduced electrochemical Ostwald ripening of the silver nanoparticles via UV irradiation[24]. This previous study showed that the ripened nanocubes exhibit enhanced plasmonic absorption after illumination.

When comparing $Cs_2AgInCl_6$ nanocubes and nanoplates, where the perovskite part has a different shape, cube vs. plate, a plasmonic feature is observed on the nanoplates but not on the nanocubes. The metallic silver surface plasmon at 2.86 eV (433 nm) is evident in the absorbance spectrum of nanoplates, indicating their different nature. This starkly contrasts with the absorption properties of nanocubes, which do not exhibit any plasmonic peak. A plausible explanation for this contrast is that the nanoplates are relatively thinner, with only a few unit cells, as opposed to the nanocubes. Consequently, most silver nanoparticles grow in a way that exposes their surfaces to a lower dielectric environment, enhancing the plasmonic effect.

Figure 3(b) shows the photoluminescence (PL) spectra of the nanoplates in comparison to nanocubes, dispersed in hexane. The nanoplates exhibit a broad and asymmetric PL spectrum at around 2.05eV (excited at 4.35eV). This asymmetric spectrum can be well fitted by using a double Gaussian function, giving rise to two emission bands centered at 1.83 eV (full width at half maximum (FWHM) = 0.63 eV), and at 2.27eV (FWHM = 0.72 eV). According to previous studies [12,36], these two emission bands arise from self-trapped excitons (STEs) and free excitons (FEs), respectively [36]. Our fits show that the ratio between STEs/ FEs amplitude is two times larger in nanoplates than in nanocubes (0.72:0.38). Although small, there is a statistical increase of the $Cs_2AgInCl_6$ nanoplates quantum yield of 0.97%±0.00(2), relatively to negligible quantum yield of nanocubes, 0.24%±0.00(1) (average

PLQY, as measured on different nanoplates and nanocubes, see Tables S1-S2). We hypothesize two mechanisms that involve symmetry breaking and may contribute to this effect. The first mechanism is the nanoplate's thin thickness, which may lead to increased strain in the nanoparticles and an increased Jahn‑Teller distortion of the $[AgCl_6]^{5-}$ octahedra [12,14,15,17,45,46], leading to an increased STEs mechanism (see Figure S7). The second mechanism involves a spatially extended silver plasmon-induced field, that may break the local symmetry of the STE and relax the constraints on the optical transition[47,48].

Contrary to very emissive CdS nanocrystals that upon the growth of a metallic tip show severe quenching of emission due to charge transfer[49,50], in our case, the radiative process is forbidden due to symmetry consideration, and thus nonradiative processes are emphasized, leading to intrinsic low PLQY, and in this case, the nanoplates provide a pathway to breaking the symmetry and allowing more radiative recombination[45]. We believe this effect will be more significant after alloying and doping $Cs_2AgInCl_6$ nanoplates, which will also increase their PLQY to significantly higher values, as happens in nanocubes.

The nanoplate's emission was stable for extended periods. After 5 months of ambient storage of one of the samples, with 1.37% PLQY as measured after the synthesis, the colloidal PLQY was measured again, showing values of 1.24%, this very negligible decrease, indicating the electronic stability of the nanoplates. TEM samples that were prepared from this solution after five months and tested in the TEM did not show a significant change in perovskite nanoplates morphology and had a clear presence of the silver decorations; indicating structural and colloidal stability of the nanocrystals over time.

The emission-excitation map of the nanoplates is shown in Figure S8(a), which clearly shows the asymmetric PLE spectra, and strengthens the claim that it combines two emission mechanisms, as mentioned before. Figure S8(b) shows the photoluminescence excitation (PLE) spectra of $Cs_2AgInCl_6$ nanoplates (blue lines) in comparison to nanocubes (purple line), where the light was collected at 600 nm (2.07 eV), and for nanocubes at 605 nm (2.05 eV). The PLE peak for the nanoplates is blue-shifted in comparison to the nanocubes. We also measured the time-resolved emission spectra (TRES) for nanoplates, the resulting value is in agreement with the long lifetime, typical for this material (see Figure S8(c)) [12,14,15,17,45,46].

To justify the claim of a spatially extended plasmon that may induce synergistic properties in the perovskite, we set out to physically measure the extent of plasmonic interaction. This challenging experiment used electron energy loss spectroscopy (EELS) to map out the plasmon interaction volume. It is important to remember that EELS measurements in perovskite nanoparticles are hard to conduct due to material sensitivity to the electron beam. An accompanying challenge is electron beam damage to the organic surface ligands, resulting in amorphous carbon contamination. Indeed, consistent physical values of lead-halide perovskite nanocrystals may be obtained using EELS[51,52]. As far as we know, this is the first EELS study for double-perovskite nanocrystals.

The EELS measurements were performed on a typical $Cs_2AgInCl_6$ nanoplate structure, as shown in TEM micrograph in Figure S9. Three regions were explored, (1) a silver nanoparticle on a perovskite matrix, (2) a mixed region containing a silver particle and perovskite, and (3) a pure perovskite region a few nanometers away from the silver nanoparticle. The EELS results, showcasing both the measured $Cs_2AgInCl_6$ bandgap and the metallic silver surface plasmon peak. The most notable finding is observing the plasmon peak extending beyond the silver nanoparticle by several nanometers, with its intensity gradually diminishing in areas (2) and (3) yet remaining visibly distinct. These results fall in line with the electric field simulations we performed for the plasmonic delocalization of a 2-nanometer silver particle (Figure S10). Indeed, our simulation predicts that the plasmonic electric field penetrates a few nanometers into the perovskite crystal, decaying gradually and less abruptly compared to a vacuum, or a Hexane medium. These EELS measurements of the $Cs_2AgInCl_6$ two-dimensional nanoplates structure provide evidence for the presence of a localized surface plasmon resonance.

In order to study the effect of prolonged irradiation and the stability of the nanoplate structures' optical properties, the product solution was placed in a quartz cuvette. It was excited for 1 hour at a wavelength of 300 nm by a Xenon lamp. Six PL spectra were measured during this time; see Figure 3c. The PL intensity was consistently reduced during the experiment, and the quantum yield was reduced from 1.24% to 0.9%. In a different experiment, the absorption spectrum of the silver-nanoplate hybrid structure was measured for 15.5 hours of UV lamp irradiation (254nm (488eV), 5 mW), see Figure 3d. The intensity of the silver plasmon peak was consistently enhanced during the experiment. A slight blue shift from 432nm to 425nm after 15.5 hours of irradiation was observed. We assign this to a change in the dielectric environment of the surrounding silver decorations inducing a plasmon resonance shift (see Extended Explanations -S1 in SI). The above hypothesis is supported by TEM characterization of the irradiated sample (after 15.5 hours of irradiation) displaying large silver particles probably due to photochemically activated Ostwald ripening (Figure 3e)[24]. We observe that bigger silver nanoparticles (estimated sizes) expose more silver interfaces with the hexane solvent than interfaces embedded in the perovskite. This phenomenon manifests in a blue-shifted plasmonic peak.

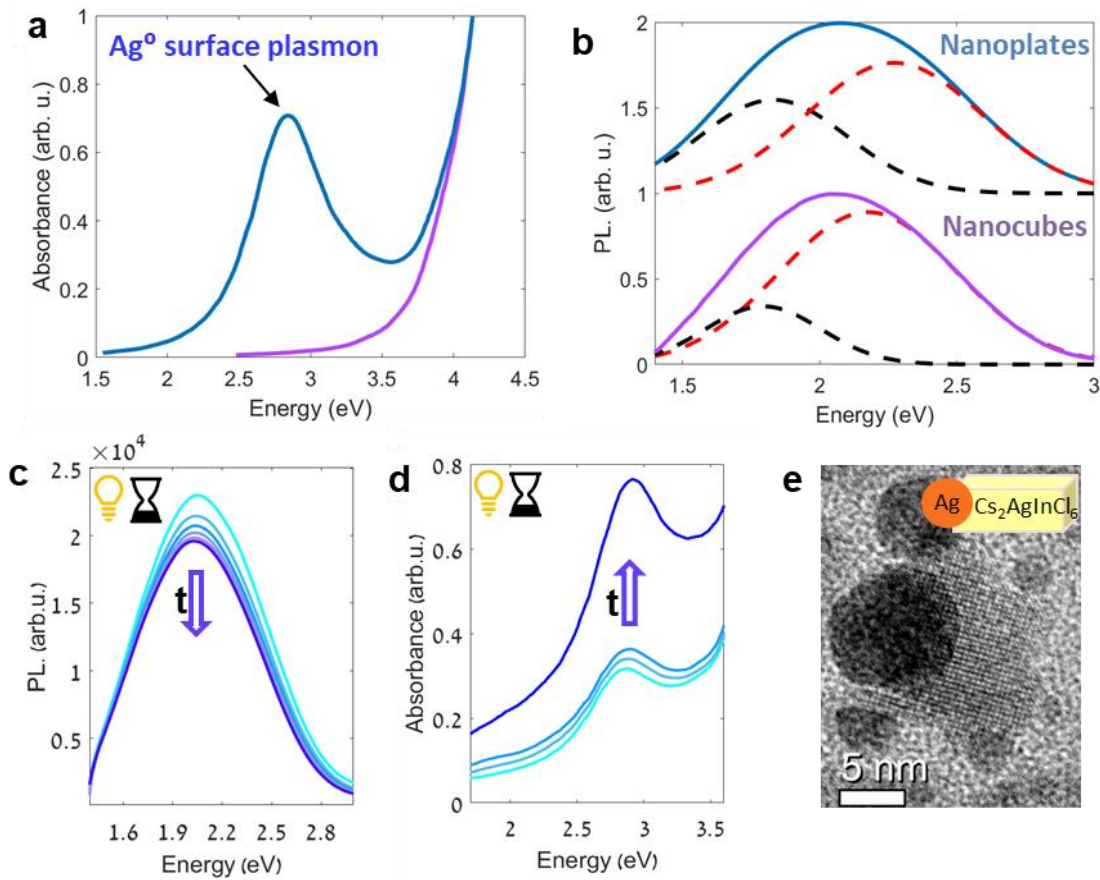

Figure 3: (a) and (b) showing the absorbance and photoluminescence (PL) spectra of $Cs_2AgInCl_6$ nanoplates (blue lines) in comparison to $Cs_2AgInCl_6$ nanocubes (purple lines). The absorbance spectrum of $Cs_2AgInCl_6$ nanoplates clearly shows the metallic silver surface plasmon due to nanoplates' small thickness compared to nanocubes. The PL spectra include their Gaussian fits to STEs (black fits) and FEs (red fits) mechanisms, respectively. The ratio between STEs/ FEs amplitude is two times larger in nanoplates than in nanocubes (0.72:0.38). The nanoplates were excited at a wavelength of 285 nm (4.35 eV), and the nanocubes were excited at a wavelength of 300 nm (4.13 eV).

(c) Optical absorbance measurements of $Cs_2AgInCl_6$ two-dimensional nanoplates before and after (15 min, 70 min, 15.5 hours) UV irradiation (254 nm (4.88 eV), 5 mW). The plasmon is blue-shifted along the experiment (from 432nm (2.87 eV) before the irradiation to 430nm (2.88 eV), 427nm (2.90 eV), and finally 425nm (2.92 eV)).

(d) PL spectrum of $Cs_2AgInCl_6$ two-dimensional nanoplates before and after UV irradiation (excited at a wavelength of 300 nm(4.13 eV)). PL intensity decreases after irradiation (10min, 20min, 30 min, 40 min, 50min, 60 min). (e) A TEM micrograph of typical $Cs_2AgInCl_6$ two-dimensional nanoplates after long UV irradiation (15.5 hours, 254 nm (4.88 eV), 5 mW) and inset of its cartoon.

It is interesting to note that the perovskite nanoplates and the silver nanoparticles tend to have almost epitaxial relationships between them (preferred d-spaces of 2.6Å {400} and 2.4Å {111} for the perovskite and the silver, respectively (see Extended Explanations- S2 in SI). When comparing the stability of nanocubes to nanoplates on different surfaces, it was found that nanoplates are less stable due to their larger surface-to-volume ratio and tendency to stack and coalesce (see Figure S11 and Figure S13). This instability is exacerbated in $Cs_2AgInCl_6$ two-dimensional nanoplate materials due to the presence of un-passivated silver islands on the surface. Surprisingly, this instability results in an interesting and potentially useful phenomenon - after a few days of storage at room temperature, small nanoplates spontaneously transform into larger lateral nanosheets on the substrate. These nanoplates were deposited on both a silicon substrate and a carbon film on a copper TEM grid and after a few days of storage, were characterized using a TEM (Figure 4a-b), a Scanning Electron Microscope (SEM) (Figure 4c), and Atomic Force Microscope (AFM) (Figure 4d). Both experiments showed that the nanoplates fused together to form nanosheets with larger lateral dimensions and rectangular shapes. The height profiles of the nanosheets were uniform at ~6 nanometers (Figure 4e), which is similar to the thickness of the small nanoplates, indicating that the smaller nanoplates oriented themselves and fused to form larger sheets (see Figure S12). These results suggest that nanosheets are thermodynamically more stable than nanoplates due to their favorable area-to-volume ratio, and that their preferred thickness is 2 unit cells.

We see clear evidence indicating that metallic silver decorations are also found on these larger fused sheets, which can grow to sizes of several microns, see an extension (Figure S13 and Extended Explanations (S2) in SI). Previous research has shown that the plasmonic nature of these silver islands can be used for catalysis [44]. For instance, $Cs_2AgInCl_6$ has a large band-gap (3.23-3.53eV[23]) which makes it transparent to visible light and opens up possibilities for its application on glass windows to provide additional functionality. For example, harnessing UV from sunlight for decomposing airborne pollutants. One possible implication of the fussing tendency of nanoplates is that in the future, it may be possible to spontaneously create extended thin films on various surfaces simply by spray-coating techniques. However, this is a long-term goal that will require further research to achieve.

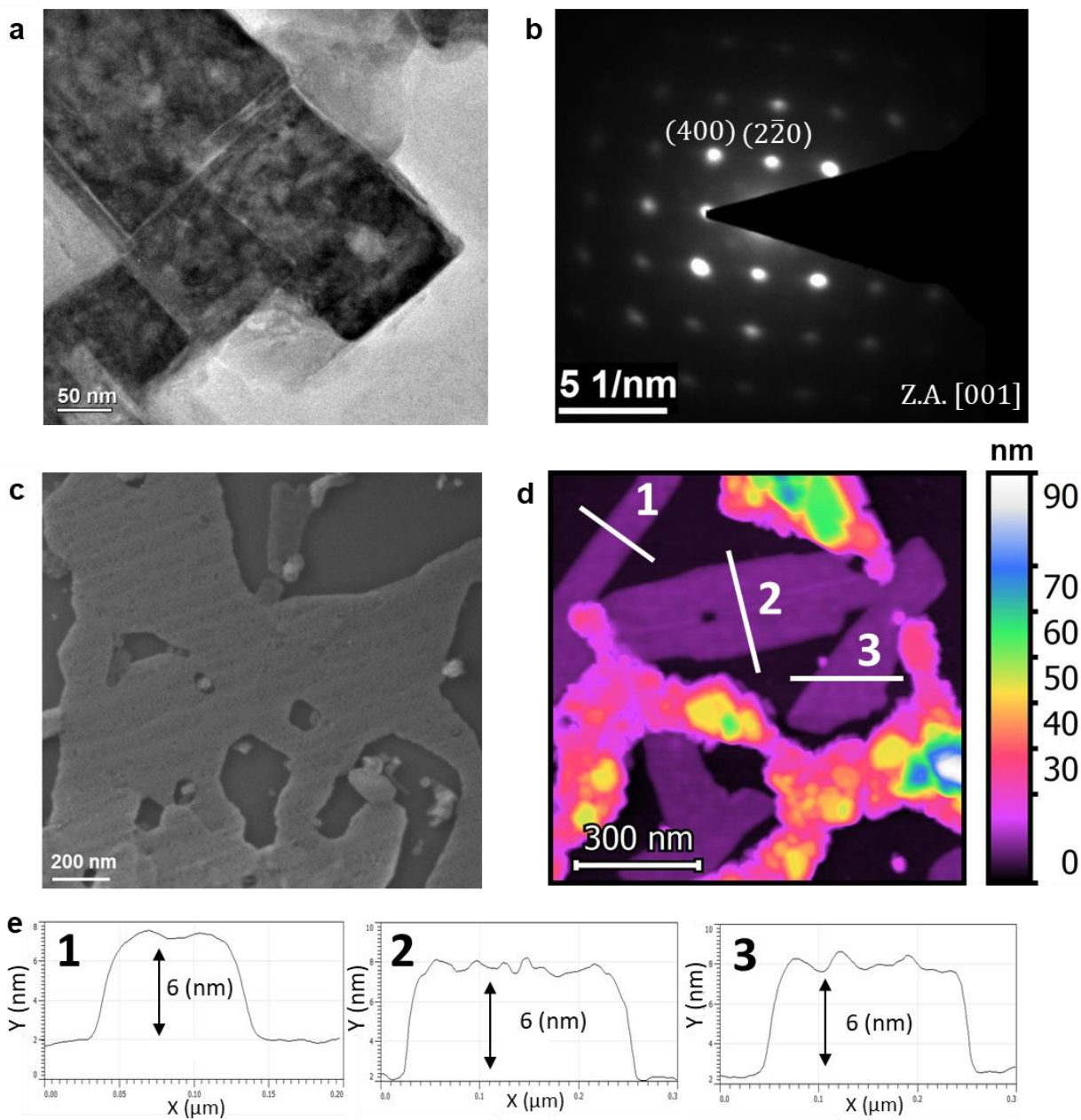

Figure 4: (a) TEM micrograph was taken from a typical area of $Cs_2AgInCl_6$ nanoplates sample, after a few days of storage at room temperature, showing the nanoplates fussing to a big rectangular nanoplate on the TEM grid. (b) Selected area diffraction is taken from (a), confirming a single big nanosheet matching [001] zone axis of $Cs_2AgInCl_6$ (ICDD 01-085-7533). (c) SEM micrograph and (d) AFM image was taken from a typical area of $Cs_2AgInCl_6$ nanoplates sample, after a few days of storage at room temperature, showing the nanoplates fussing to a big nanosheet on a silicon substrate. (e) The height profiles of (d) showcase uniform ~6 nanometers thickness (with ligands).[23]

In this work, we developed a novel colloidal synthesis for $Cs_2AgInCl_6$ two-dimensional nanoplates with metallic silver decorations. These nanoplate structures show optoelectronic properties modified by the plasmonic properties of the metallic silver nanoparticle decorations. The transformation of the nanoplates into larger thin sheets upon deposition on a substrate further highlights their potential as building blocks for functional surfaces. Future research in this area should focus on precise control of these nanoparticles, investigating the influence of other factors such as doping and alloying with other elements, and exploring their potential applications in optoelectronic devices. The results of this study pave the way for the development of new and exciting nanoplates semiconductor-metal nanoparticle systems with enhanced properties for a wide range of applications.

## Supporting Information

Supporting Information is available free of charge:

1. Materials and Methods.

2. Supplementary Text (Extended Explanations): S1-S3

3. Figures: Figure S1 to Figure S13.

4. Tables: Table S1 to Table S2.

## Author Information

**Corresponding Author**

**Yehonadav Bekenstein** − Department of Materials Science and Engineering, The Solid-State Institute, and The Nancy and Stephen Grand Technion Energy Program, Technion – Israel Institute of Technology, 32000 Haifa, Israel ; orcid.org/ 0000-0001-6230-5182;

Email: bekenstein@technion.ac.il

**Authors**

**ORCID** 

**Sasha Khalfin:** Department of Materials Science and Engineering, Technion − Israel Institute of Technology, 32000 Haifa, Israel; orcid.org/0000-0003-2983-3367

**Noam Veber:** Department of Materials Science and Engineering and the Solid-State Institute, Technion − Israel Institute of Technology, 32000 Haifa, Israel; orcid.org/0000-0001-7513-6357

**Saar Shaek:** Department of Materials Science and Engineering, Technion − Israel Institute of Technology, 3200003 Haifa, Israel; orcid.org/0000-0001-9599-6098

**Betty Shamaev:** Department of Materials Science and Engineering, Technion − Israel Institute of Technology, 32000 Haifa, Israel; orcid.org/0009-0005-3679-2825

**Shai Levy:** Department of Materials Science and Engineering, Technion − Israel Institute of Technology, 3200003 Haifa, Israel; orcid.org/0000-0001-6376-0486

**Shaked Dror:** Department of Materials Science and Engineering, Technion − Israel Institute of Technology, 32000 Haifa, Israel; orcid.org/0000-0002-6273-7054

**Yaron Kauffmann:** Department of Materials Science and Engineering, Technion − Israel Institute of Technology, 32000 Haifa, Israel; orcid.org/0000-0002-0117-6222

**Maria Koifman Khristosov:** Department of Materials Science and Engineering, Technion − Israel Institute of Technology, 32000 Haifa, Israel; orcid.org/ 0000-0003-0577-6572

## Data Availability

The data supporting this article have been included as part of the Supplementary Information.

## Acknowledgments

We gratefully thank the scientific staff of our microscopy center (MIKA), Dr. Michael Kalina, Dr. Galit Atiya, and Dr. Ellina Kesselman for their continuous support and helpful advice. YB thanks the Nancy and Stephen Grand Technion Energy Program for their generous support and support of the Technion Russel Berrie Nanotechnology Institute, and of the Technion Helen Diller Quantum Center.

**Funding:** This work is supported by the Israel Science Foundation grant number 890015 and by the European Union's Horizon 2020 research and innovation program under grant agreement No. 949682-ERC-HeteroPlates.

**Competing interests:** The authors declare no competing financial interests.

# References

[1]     Y. Bekenstein, B. A. Koscher, S. W. Eaton, P. Yang, A. P. Alivisatos, *J Am Chem Soc* **2015**, *137*, 16008.

[2]     L. Protesescu, S. Yakunin, M. I. Bodnarchuk, F. Krieg, R. Caputo, C. H. Hendon, R. X. Yang, A. Walsh, M. V Kovalenko, *Nano Lett* **2015**, *15*, 3692.

[3]     Z. Liu, Y. Bekenstein, X. Ye, S. C. Nguyen, J. Swabeck, D. Zhang, S. T. Lee, P. Yang, W. Ma, A. P. Alivisatos, *J Am Chem Soc* **2017**, *139*, 5309.

[4]     E. Cohen, A. Nagel, M. Fouchier, L. Popilevsky, Y. Kauffmann, S. Khalfin, S. Dror, Y. Bekenstein, *Chemistry of Materials* **2022**, *34*, 5377.

[5]     N. Veber, B. Shamaev, S. Shaek, E. H. Massasa, J. Zimmerman, O. Be'er, S. Levy, Y. Bekenstein, *Adv Opt Mater* **2024**.

[6]     M. V Kovalenko, L. Protesescu, M. I. Bodnarchuk, *Science (1979)* **2017**, *358*, 745.

[7]     Q. A. Akkerman, G. Rainò, M. V. Kovalenko, L. Manna, *Nat Mater* **2018**, *17*, 394.

[8]     S. Levy, O. Be'er, N. Veber, C. Monachon, Y. Bekenstein, *Nano Lett* **2023**.

[9]     M. R. Filip, F. Giustino, *Journal of Physical Chemistry C* **2016**, *120*, 166.

[10]    E. T. McClure, M. R. Ball, W. Windl, P. M. Woodward, *Chemistry of Materials* **2016**, *28*, 1348.

[11]    M. R. Filip, S. Hillman, A. A. Haghighirad, H. J. Snaith, F. Giustino, *Journal of Physical Chemistry Letters* **2016**, *7*, 2579.

[12]    J. Luo, X. Wang, S. Li, J. Liu, Y. Guo, G. Niu, L. Yao, Y. Fu, L. Gao, Q. Dong, C. Zhao, M. Leng, F. Ma, W. Liang, L. Wang, S. Jin, J. Han, L. Zhang, J. Etheridge, J. Wang, Y. Yan, E. H. Sargent, J. Tang, Vol. 563, Nature, **2018**, pp. 541–545.

[13]    S. Dror, S. Khalfin, N. Veber, A. Lang, Y. Kauffmann, M. Koifman Khristosov, R. Shechter, B. Pokroy, I. E. Castelli, Y. Bekenstein, *Chemistry of Materials* **2023**.

[14]    F. Locardi, M. Cirignano, D. Baranov, Z. Dang, M. Prato, F. Drago, M. Ferretti, V. Pinchetti, M. Fanciulli, S. Brovelli, L. De Trizio, L. Manna, *J Am Chem Soc* **2018**, *140*, 12989.

[15]    J. C. Dahl, W. T. Osowiecki, Y. Cai, J. K. Swabeck, Y. Bekenstein, M. Asta, E. M. Chan, A. P. Alivisatos, *Chemistry of Materials* **2019**, *31*, 3134.

[16]    F. Zhao, Z. Song, J. Zhao, Q. Liu, *Inorg Chem Front* **2019**, *6*, 3621.

[17]    Y. Liu, Y. Jing, J. Zhao, Q. Liu, Z. Xia, *Chemistry of Materials* **2019**, *31*, 3333.

[18]    Y. Liu, A. Nag, L. Manna, Z. Xia, *Angewandte Chemie International Edition* **2021**, *60*, 11592.

[19]    L. Protesescu, S. Yakunin, M. I. Bodnarchuk, F. Krieg, R. Caputo, C. H. Hendon, R. Xi Yang, A. Walsh, M. V. Kovalenko, *Nano Lett* **2015**, *15*, 3692.

[20]    D. Zhang, S. W. Eaton, Y. Yu, L. Dou, P. Yang, *J Am Chem Soc* **2015**, *137*, 9230.

[21]    Y. Bekenstein, B. A. Koscher, S. W. Eaton, P. Yang, A. P. Alivisatos, *J Am Chem Soc* **2015**, *137*, 16008.

[22]     E. H. Massasa, R. Strassberg, A. Vurgaft, Y. Kauffmann, N. Cohen, Y. Bekenstein, *Nano Lett* **2021**, *21*, 5564.

[23]     S. Khalfin, Y. Bekenstein, Vol. 11, Royal Society of Chemistry, **2019**, pp. 8665–8679.

[24]     S. Levy, S. Khalfin, N. G. Pavlopoulos, Y. Kauffmann, G. Atiya, S. Shaek, S. Dror, R. Shechter, Y. Bekenstein, *Chemistry of Materials* **2021**, *33, 7*, 2370.

[25]     S. Khalfin, N. Veber, S. Dror, R. Shechter, S. Shaek, S. Levy, Y. Kauffmann, L. Klinger, E. Rabkin, Y. Bekenstein, *Adv Funct Mater* **2022**, *32*, 2110421.

[26]     S. Shaek, S. Khalfin, E. H. Massasa, A. Lang, S. Levy, L. T. J. Kortstee, B. Shamaev, S. Dror, R. Lifer, R. Shechter, Y. Kauffmann, R. Strassberg, I. Polishchuk, A. B. Wong, B. Pokroy, I. E. Castelli, Y. Bekenstein, *Chemistry of Materials* **2023**, *35*, 9064.

[27]     U. Banin, Y. Ben-Shahar, K. Vinokurov, Vol. 26, **2014**, pp. 97–110.

[28]     Y. Ben-Shahar, D. Stone, U. Banin, *Chemical Reviews* **2023**, *123, 7*, 3790.

[29]     Y. Wang, H. N. Nyiera, A. W. Mureithi, Y. Sun, T. Mani, J. Zhao, *Journal of Physical Chemistry C* **2022**.

[30]     Y. Ben-Shahar, J. P. Philbin, F. Scotognella, L. Ganzer, G. Cerullo, E. Rabani, U. Banin, *Nano Lett* **2018**, *18*, 5211.

[31]     B. J. Roman, J. Otto, C. Galik, R. Downing, M. Sheldon, *Nano Lett* **2017**, *17*, 5561.

[32]     S. K. Balakrishnan, P. V Kamat, **2023**, *13*, 24.

[33]     M. Gong, M. Alamri, D. Ewing, S. M. Sadeghi, J. Z. Wu, M. Gong, M. Alamri, J. Z. Wu, D. Ewing, S. M. Sadeghi, *Advanced Materials* **2020**, *32*, 2002163.

[34]     F. A. Rodríguez Ortiz, B. J. Roman, J. R. Wen, N. Mireles Villegas, D. F. Dacres, M. T. Sheldon, *Nanoscale* **2019**, *11*, 18109.

[35]     S. Chen, D. Lyu, T. Ling, W. Guo, R. Li, / Chemcomm, C. Communication, *Chemical Communications* **2018**, *54*, 4605.

[36]     W. Ma, H. Wu, X. Liu, R. Miao, T. Liang, J. Fan, *Adv Opt Mater* **2022**, *10*, 2200837.

[37]     T. Bai, B. Yang, J. Chen, D. Zheng, Z. Tang, X. Wang, Y. Zhao, R. Lu, K. Han, T. Bai, D. Zheng, Z. Tang, X. Wang, K. Han, B. Yang, J. Chen, Y. Zhao, R. Lu, *Advanced Materials* **2021**, *33*, 2007215.

[38]     H. Tang, Y. Xu, X. Hu, Q. Hu, T. Chen, W. Jiang, L. Wang, W. Jiang, *Advanced Science* **2021**, *8*, 2004118.

[39]     S. E. Creutz, E. N. Crites, M. C. De Siena, D. R. Gamelin, *Nano Lett* **2018**, *18*, 1118.

[40]     Y. Zhang, T. D. Siegler, C. J. Thomas, M. K. Abney, T. Shah, A. De Gorostiza, R. M. Greene, B. A. Korgel, *Chemistry of Materials* **2020**.

[41]     Y. Bekenstein, J. C. Dahl, J. Huang, W. T. Osowiecki, J. K. Swabeck, E. M. Chan, P. Yang, A. P. Alivisatos, *Nano Lett* **2018**, *18*, 3502.

[42]     K. L. Kelly, E. Coronado, L. L. Zhao, G. C. Schatz, *Journal of Physical Chemistry B* **2003**, *107*, 668.

[43]   J. J. Mock, M. Barbic, D. R. Smith, D. A. Schultz, S. Schultz, *J Chem Phys* **2002**, *116*, 6755.

[44]   A. L. Koh, K. Bao, I. Khan, W. E. Smith, G. Kothleitner, P. Nordlander, S. A. Maier, D. W. Mccomb, *ACS Nano* **2009**, *3*, 3015.

[45]   F. Locardi, E. Sartori, J. Buha, J. Zito, M. Prato, V. Pinchetti, M. L. Zaffalon, M. Ferretti, S. Brovelli, I. Infante, L. De Trizio, L. Manna, *ACS Energy Lett* **2019**, *4*, 1976.

[46]   Y. Mahor, W. J. Mir, A. Nag, *Journal of Physical Chemistry C* **2019**.

[47]   D. Weinberg, Y. Park, D. T. Limmer, E. Rabani, **2023**.

[48]   P. K. Jain, D. Ghosh, R. Baer, E. Rabani, A. P. Alivisatos, *Proc Natl Acad Sci U S A* **2012**, *109*, 8016.

[49]   P. Rukenstein, A. Teitelboim, M. Volokh, M. Diab, D. Oron, T. Mokari, *J. Phys. Chem. C* **2016**, *120*.

[50]   A. E. Saunders, I. Popov, U. Banin, **2006**.

[51]   B. Mitchell, E. Herrmann, J. Lin, L. Gomez, C. De Weerd, Y. Fujiwara, K. Suenaga, T. Gregorkiewicz, *Eur J Phys* **2018**, *39*, 055501.

[52]   J. Lin, L. Gomez, C. De Weerd, Y. Fujiwara, T. Gregorkiewicz, K. Suenaga, *Nano Lett* **2016**, *16*, 7198.

TOC:

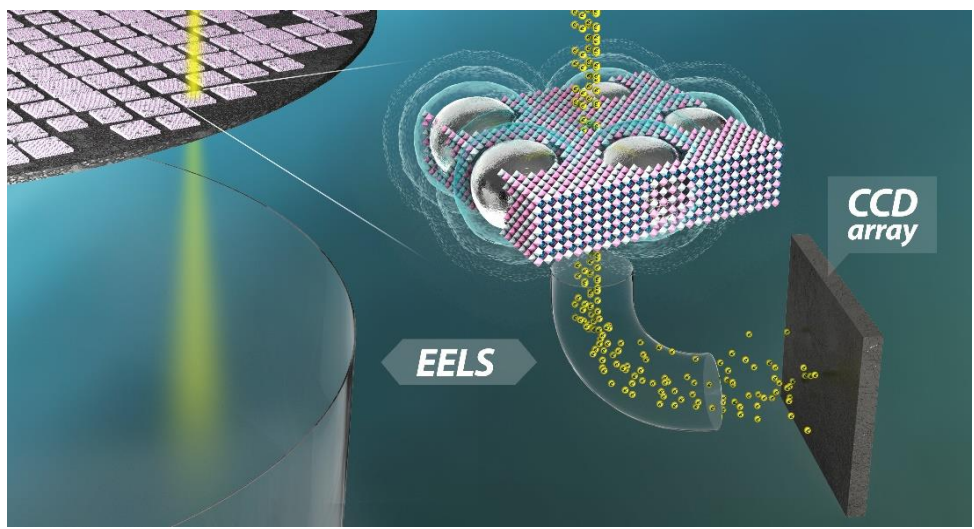

This study presents a novel colloidal synthesis of a $Cs_2AgInCl_6$ perovskite two-dimensional nanoplates with small lateral dimensions and metallic decorations; and compares those nanoplates to the known $Cs_2AgInCl_6$ perovskite nanocubes.

# Supporting Information

# Synthesis and Characterization of Two-dimensional $Cs_2AgInCl_6$ Nanoplates as Building Blocks for Functional Surfaces

*Sasha Khalfin[#1], Noam Veber[#1,2], Saar Shaek[1], Betty Shamaev[1,2], Shai Levy[1], Shaked Dror[1], Yaron Kauffmann[1], Maria Koifman Khristosov[1], and Yehonadav Bekenstein[1,2,3]*

1- Department of Materials Science and Engineering, Technion – Israel Institute of Technology, 3200003 Haifa, Israel.

2- The Solid-state institute, Technion – Israel Institute of Technology, 3200003 Haifa, Israel.

3- The Nancy and Stephen Grand Technion Energy Program, Technion – Israel Institute of Technology, 32000 Haifa, Israel.

*Corresponding author. Email: bekenstein@technion.ac.il
#These authors contributed equally

**Materials and Methods:**

*Materials*: Benzoyl bromide (97%, Aldrich), Cesium acetate (Cs(OAc), 99.9%, Aldrich), chlorotrimethylsilane (TMSCl, ≥99%, Sigma-Aldrich), hexane (99.9%, Fisher Scientific), indium (III) acetate (In(OAc)$_3$, 99.99%, Sigma-Aldrich), octadecene (ODE, 90%, Sigma-Aldrich), oleic acid (OA, ≥99%, Sigma-Aldrich), oleylamine (OLA, 70% or ≥98%, Sigma-Aldrich), o-xylene anhydrous (97%, Aldrich), silver acetate (Ag(OAc), 99.99%, Aldrich), toluene (A.R., Sigma-Aldrich).

*Synthesis of $Cs_2AgInCl_6$ nanoplates*: Synthesis of $Cs_2AgInCl_6$ nanoplates was performed by using a modified hot-injection method: 136.2 mg (0.71 mmol) of Cs(OAc), 41.7 mg (0.25 mmol) of Ag(OAc), 146.8 mg (0.50 mmol) of In(OAc)$_3$, OA (2.8 mL), OLA (0.66 mL) were loaded into a 3-neck flask, along with ODE (10 mL) and magnetic stirring bar. This reaction mixture was heated to 110°C under vacuum for 1 h. The reaction mixture was initially colorless and then gradually changed from light yellow to brown. After that, the solution was heated to 170 °C under a nitrogen atmosphere. TMSCl (0.4 mL) was swiftly injected when the temperature reached 170°C. The reaction solution continued to be heated to 180 °C, and after 1 min at this temperature, immediately cooled to room temperature in an ice-water bath. Then the solution was decanted into a centrifugal tube and centrifuged at 8000 rpm for 15 min. The brown supernatant is removed. The pale precipitate was dispersed in 10 mL toluene and centrifuged at 8000 rpm for 15 min. Then, the supernatant is discarded. The precipitate is re-dispersed in 10 mL hexane with sonication (1 min) and centrifuged at 8000 rpm for 15 min. After that, the colloidal NCs were obtained by discarding the bottom precipitate.

**Synthesis of $Cs_2AgInCl_6$ nanocubes**: In a typical synthesis, 25 mg (0.125 mmol) of cesium acetate, 40 mg (0.24 mmol) of silver acetate, and 80 mg (0.25 mmol) of indium (III) acetate were placed into a 20 ml glass vial with a magnetic stirring bar. O-xylene (5 mL), oleic acid (1.25mL), and oleylamine (0.375 mL) were added, and the vial was heated to 100°C for 5 min. Benzoyl chloride (0.200 mL, 1.5 mmol) was injected quickly. Then, the vials were left to cool to room temperature in a water bath. For measurements, the nanocrystal reaction mixture was precipitated by centrifugation at 9,000 rpm for 10 min and the resulting pellet was redispersed in hexane.

## Characterizations Methods:

*X-ray Diffraction:* The nanocrystal solution in hexane was dropped on a glass substrate (rectangular micro slides, 76 x 26 [mm]) and then moved to a hotplate (80°C) to evaporate the solvent. The X-ray beam focused on the resulting film. Measurements were taken using a Rigaku Smart-Lab 9 kW high-resolution X-ray diffractometer, equipped with a rotating anode X-ray source. We use a "Glancing mode" (grazing angle) method (2-theta), which is suitable for measuring thin films, with a 1.54Å (Cu Kα) wavelength. The X-rays source was fixed on ω=0.4°, and the detector moved in the range of 2θ=20°-90°.

*Transmission Electron Microscopy (TEM) and Scanning Transmission Electron Microscopy (STEM) Characterization:* One drop of dilute nanocrystal solution in hexane (1:20 dilution) was cast onto a TEM grid (amorphous carbon film on 300 mesh copper). The samples were observed in TEM mode with a Thermo Fisher/FEI Tecnai G2 T20 S-Twin

LaB6 TEM operated at 200KeV, with a 1Kx1K Gatan 694 slow-scan CCD. High-resolution imaging, diffraction patterns acquisition, and chemical mapping were done in a Thermo Fisher/FEI Titan- Themis double Cs-corrected HR-S/TEM, operated at 200kV and equipped with a Ceta2 4Kx4K camera (for TEM mode) and a Bruker Dual-X EDX detectors. EDS maps were acquired, post processed and analyzed using the Velox software (Thermo-Fisher, USA).

***Electron energy loss spectroscopy (EELS) measurements:*** 2μL of dilute nanocrystal solution in hexane (1:20 dilution) was cast onto a TEM grid (amorphous ultrathin carbon type-A film on 400 mesh copper).
For characterization at the atomic scale a Thermo-Fisher/FEI Titan Themis double Cs-Corrected HR-S/TEM operated at 200KeV was used with monochromator. The microscope is equipped with a Gatan Quantum ER965 dual-EELS detector (Gatan, USA) for electron energy loss spectroscopy (EELS) analysis. The energy spread was 160 meV. The zero-loss peak was removed by substruction of the mirrored left hand tail of the zero loss peak.

***Atomic force microscopy (AFM):*** One drop of dilute nanocrystal solution in hexane (1:20 dilution) was cast onto a silicon substrate. The AFM measurements were taken using neaspec probe in tapping mode. The resonance frequency of the tip was 70 kHz. The image was taken in 200X200 pixel resolution. No manipulation has been conducted to the AFM image other than the usual AFM flattening compensating for the tip scan.

***Scanning Electron Microscopy (SEM):*** High-resolution SEM images were taken using a SEM Zeiss Ultra-Plus HRSEM. Samples were placed at a working distance of 3mm and measured using an acceleration voltage of 1.5KV in secondary electrons mode. The samples were made by drop-casting the NPs diluted solution on a silicon substrate.

***Spectroscopic UV-Vis Absorption, Photoluminescence (PL), and Photoluminescence Excitation (PLE) measurements:*** Absorption measurements were taken using a Synergy H1 hybrid multi-mode reader with an irradiation source of a xenon lamp (Xe900). The Bio-Tek microplate spectrometer can perform high throughput optical characterization. 200ul of the dilute sample solution was inserted into a 96 wells microplate. Photoluminescence (PL), and Photoluminescence Excitation (PLE) measurements were taken using Edinburgh FLS1000 photoluminescence spectrometer. All the samples were loaded into a quartz cuvette. The nanoplates were excited at a wavelength of 285 nm, and the nanocubes were excited at a wavelength of 300 nm. The fixed emission wavelength for PLE measurements for nanoplates was 600 nm, and for nanocubes 605 nm.

*Lifetime and photoluminescence quantum yield (PLQY):*
Life-time and photoluminescence quantum yield (PLQY) characterizations were performed using the Edinburgh FLS1000 photoluminescence spectrometer. All the samples were loaded into a quartz cuvette. The lifetime measurements were performed with multi-channel scaling (MCS) mode and conducted using a variable pulse laser (VPL). The PLQY measurements were performed with an integrating sphere holder inside the spectrometer.

**Numerical Simulation Methods:**
The numerical simulation was done using the FDTD solver program "Lumerical". The size of the Ag nanoparticle was set to 2 nm, as the average size shown in the TEM results in Figure S3, and the refractive index of the metal and the surroundings were set according to literature values[1–3]. The absorption cross section was simulated and normalized to particle size and the electric fields in the near field were simulated.

**Extended Explanations:**
*S1. Blue-shifted Ag plasmonic peak after UV irradiation*
Figure 3f shows optical absorbance measurements of $Cs_2AgInCl_6$ nanoplates before and after (15 min, 70 min, 15.5 hours) UV irradiation (254 nm, 5 mW). The plasmon is blue-shifted along the experiment (from 432nm before the irradiation to 430nm (2.88 eV), 427nm (2.90 eV), and finally 425nm (2.92 eV)). The TEM micrographs display large silver particles probably due to photochemically activated Ostwald ripening during the irradiation.
This way silver nanoparticles' dielectric properties of the surrounding medium are changed from perovskite surrounding to hexane solvent, a phenomenon that manifests in the blue-shift of the silver plasmonic peak, as presented in Figure 3f. The surface plasmon peak wavelength (λ) is related to the refractive index of the surrounding medium (n) by the following expression:
(1) $\lambda^2 = \lambda_p^2(\varepsilon^\infty + 2\varepsilon_m)$

where $\lambda_p$ is the bulk metal plasmon wavelength, $\varepsilon^\infty$ is the high frequency dielectric constant, and $\varepsilon_m = n^2$) is the optical dielectric function of the medium. Substituting eq.1 with metallic silver's bulk plasmon wavelength (138 nm [4]), the high-frequency dielectric constant of 6.07 [4], and the optical dielectric function of hexane (1.89[5]). This substitution results in a plasmon peak wavelength of 433 nm (2.86 eV). This value is in strong agreement with the observed plasmonic peak in Figure 3a at 433 nm (2.86 eV), and the absorbance before irradiation of the sample in Figure 3f at 432 nm (2.87 eV).

For these calculations, we used the dielectric function of hexane solvent and not perovskite, because of the thin thickness of the perovskite, because of which the silver nanoparticles are prominent from the perovskite surface to the external hexane's solution surrounds them. We assign the blue shift after the irradiation result to the formation of bigger silver nanocrystals that are even more exposed to the hexane's solution environment than to the perovskite environment (perovskite environment of the optical dielectric function of ~4 [4] causes red shift of the plasmonic peak). The PL decrease during the irradiation experiment can be assigned to the generation of nonradiative recombination due to the coupling of the perovskite to a relatively large silver nanoparticle and the creation of a metal-semiconductor junction.

## S2. Structural compatibility between the $Cs_2AgInCl_6$ perovskite and the silver nanoparticles

The perovskite nanoplates and the silver nanoparticles tend to have almost epitaxial relationships between them (preferred d-spaces of 2.6Å {400} and 2.4Å {111} for the perovskite and the silver, respectively), see Figure S13(a). Despite this, we also observed a few cases in which this orientation does not exist, for example, see Figure S13(b) which demonstrates nanoplates fussing to big nanosheets. Additionally, from looking at Figure S11(b) it can be seen that inside the big nanosheet there is clearly one small nanoplate that not fussed in because it rotated relative to the large nanosheet. The plates have similar (220) orientation with d-space of 3.7 Å and the silver nanoparticles have (111) orientation with d-space of 2.4 Å.

## S3. Extension explanation on XRD of the $Cs_2AgInCl_6$ perovskite nanoparticles

Broad XRD peaks in nanoparticles are assigned to Scherrer broadening due to the nanoparticles' small size. The XRD diffractograms for the product of our colloidal synthesis (Figure 1c) do not present any peaks assigned to the apparent metallic Ag nanoparticles. It is possible that in the double perovskite end product, these particles are small enough (and dispersed in the product without aggregation) so that the signal diffractions from these Ag nanoparticles do not appear in the diffractogram. Such broad peaks could be accounted for as background noise in the XRD and subtracted from the pattern during the measurement and analysis. The identity of the Ag nanoparticles was confirmed by EDX elemental mapping.

**Supporting Information Results:**

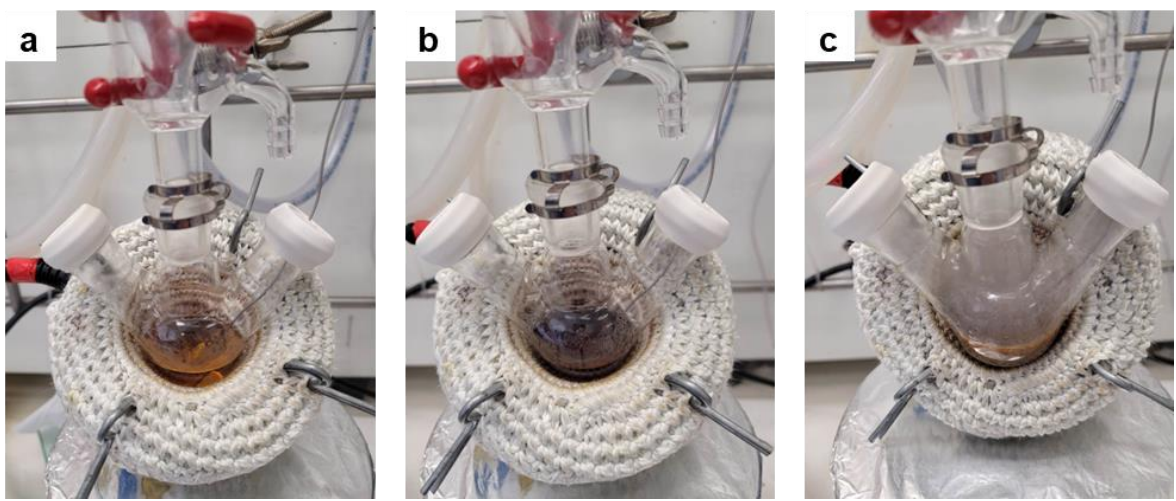

Figure S1: photos of the synthesis procedure for the $Cs_2AgInCl_6$ double perovskite nanoplates. (a) after the addition of acetates, organic ligands, ODE solvent, and starting of heating to 110°C under vacuum. (b) after 1 hour at 110°C under vacuum, and (c) after TMSCl was injected.

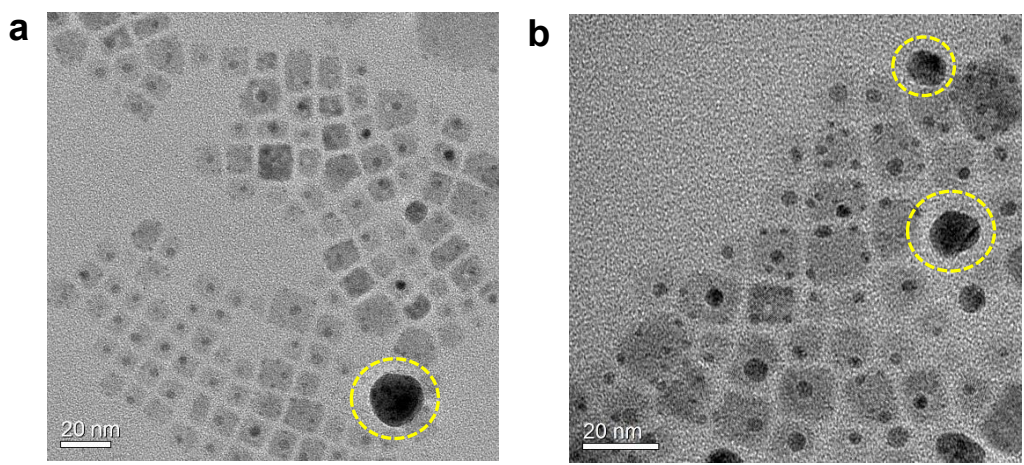

Figure S2: (a) and (b) Typical TEM micrograph of $Cs_2AgInCl_6$ nanocubes and AgCl by-product (marked by a yellow circle).

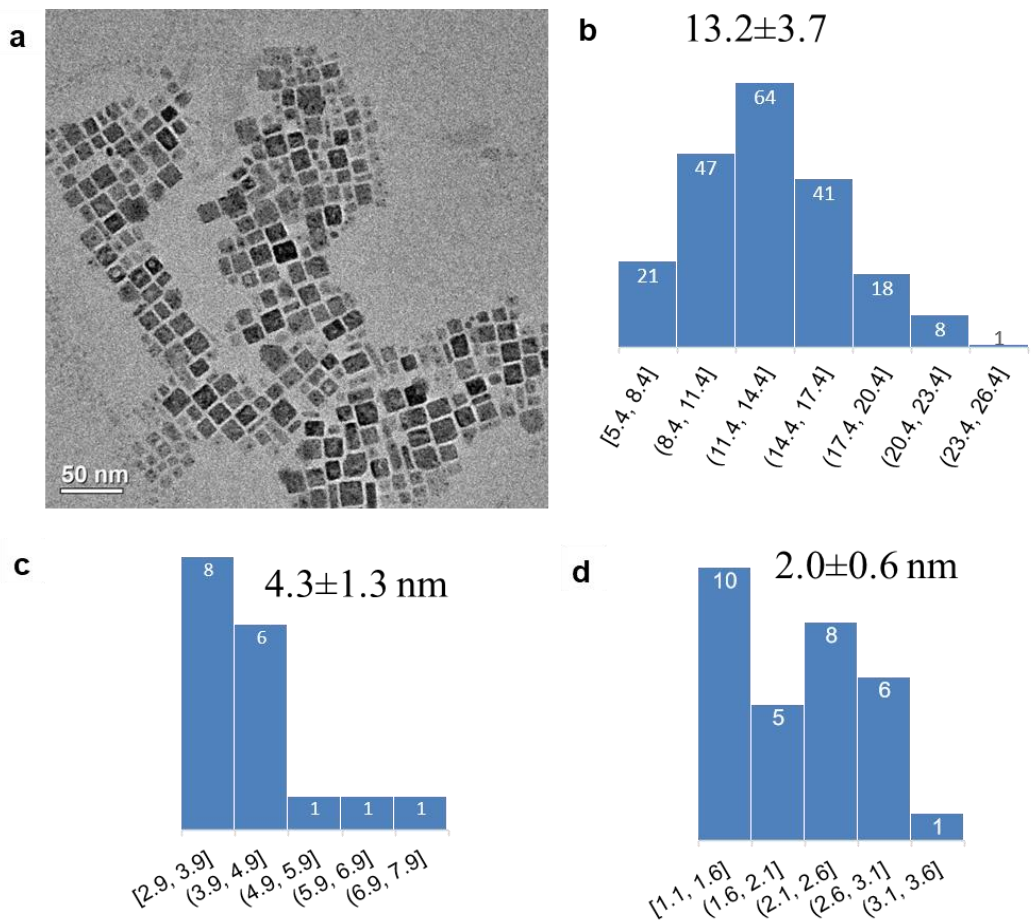

Figure S3: (a) Typical TEM micrograph of Cs$_2$AgInCl$_6$ nanoplates and histograms of its (b) lateral dimensions (13.2±3.7) nm and (c) thickness dimensions (4.3±1.3 nm).
The Cs$_2$AgInCl$_6$ nanoplates decorated by high-contrast spherical metallic silver nanoparticles that's lateral dimensions shown in (d) histogram (2.0±0.6).

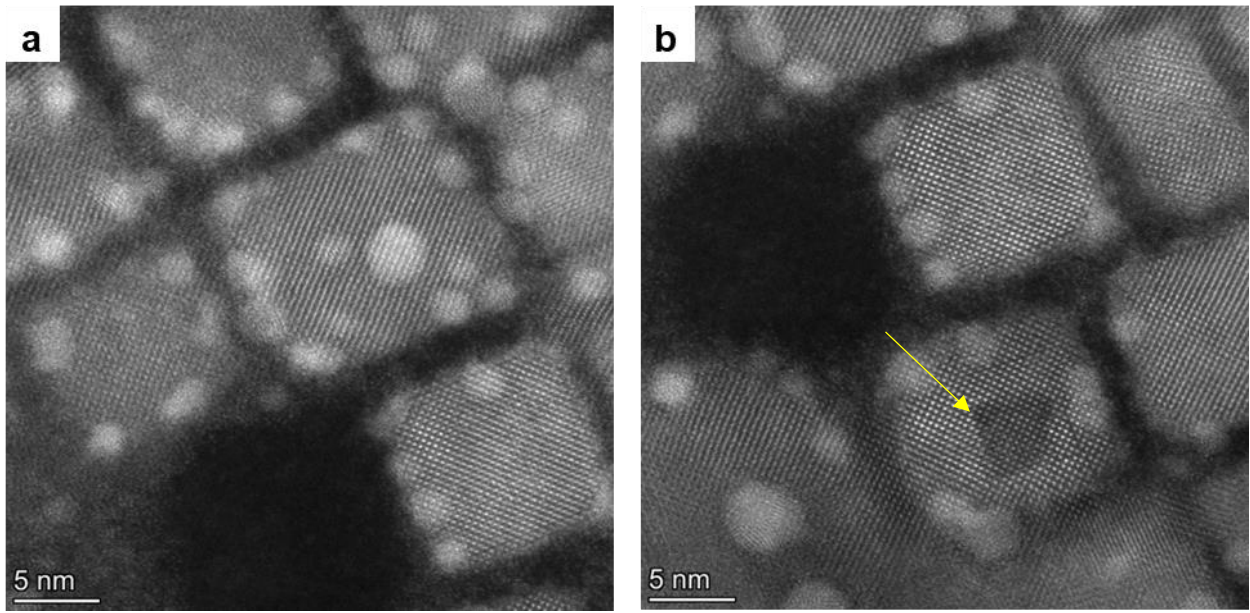

Figure S4: (a) (c) HR HAADF-STEM micrograph of $Cs_2AgInCl_6$ nanoplates. (d) Void formation inside the $Cs_2AgInCl_6$ nanoplates.

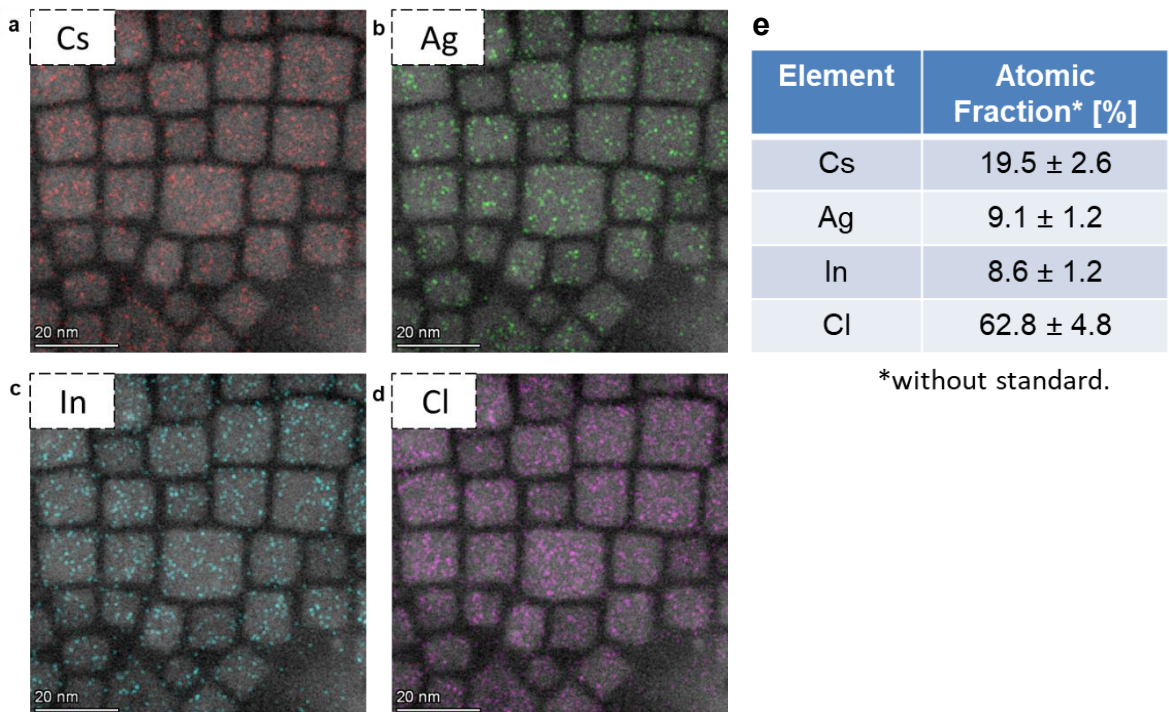

| Element | Atomic Fraction* [%] |
|---|---|
| Cs | 19.5 ± 2.6 |
| Ag | 9.1 ± 1.2 |
| In | 8.6 ± 1.2 |
| Cl | 62.8 ± 4.8 |

*without standard.

Figure S5: (a-d) EDX elemental mapping of a typical area of $Cs_2AgInCl_6$ nanoplates (analyzed only on pure perovskite area) for cesium, silver, indium, and chlorine, respectively, and (e) the resulting ratio between the elements proves the $Cs_2AgInCl_6$ composition.

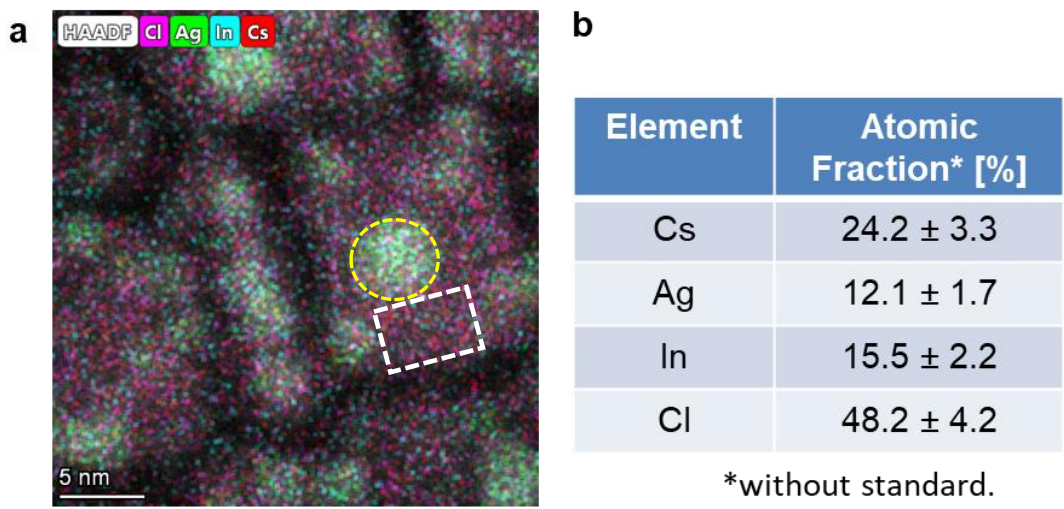

Figure S6 (a-b): EDX elemental mapping of a typical area (as marked by a white rectangle) analysis of $Cs_2AgInCl_6$ nanoplates from Figure 2(c-f) for cesium, silver, indium, and chlorine, respectively. The expected ratio of perovskite elements is achieved, with a characteristic lack of halides, previously observed in perovskites. According to the EDS, the area marked by a yellow circle is richer in silver, and this is because this is an area of the perovskite that contains a metallic nanoparticle of silver (the perovskite-silver hybrid area).

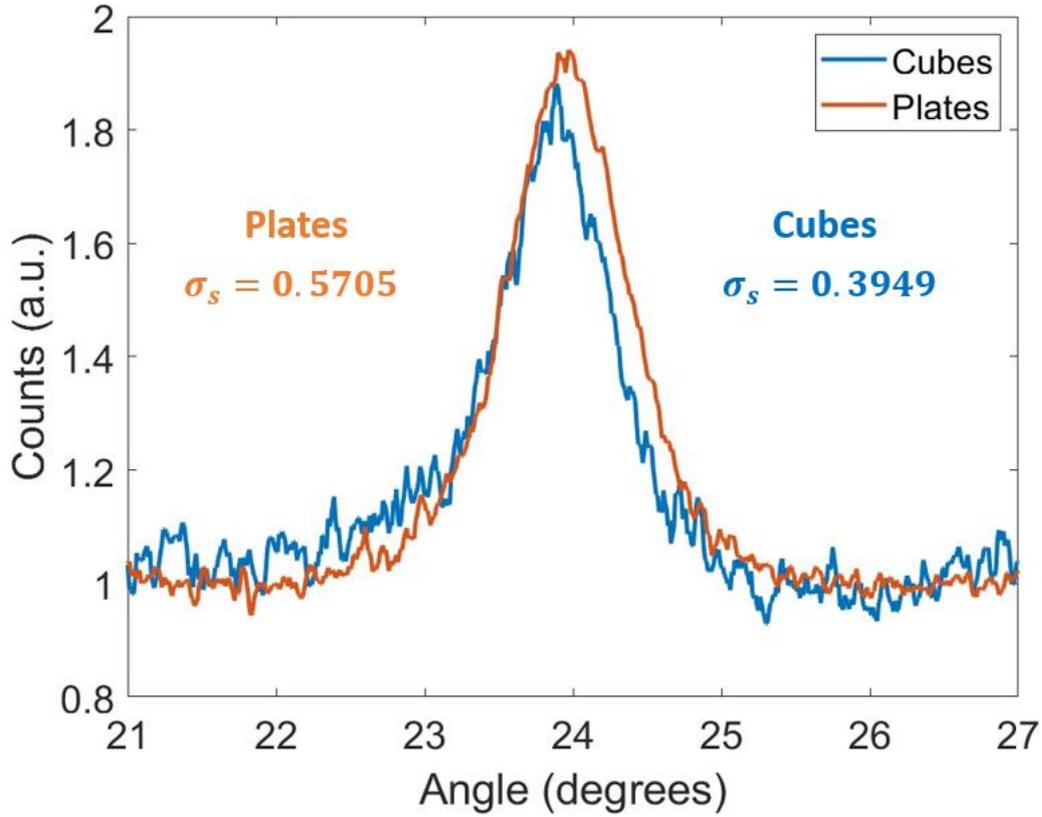

Figure S7: XRD analysis of the average micro-strain for both nanocubes and nanoplates, showing a significantly larger micro-strain for the nanoplates. Analysis is performed by fitting a Voigt function and extracting the Gaussian part: $\sigma_s = \frac{W_G}{4\sqrt{2\ln 2}\tan\theta_B}$ where $W_G$ is the Gaussian width, and $\theta_B$ is the Bragg's angle.[6]

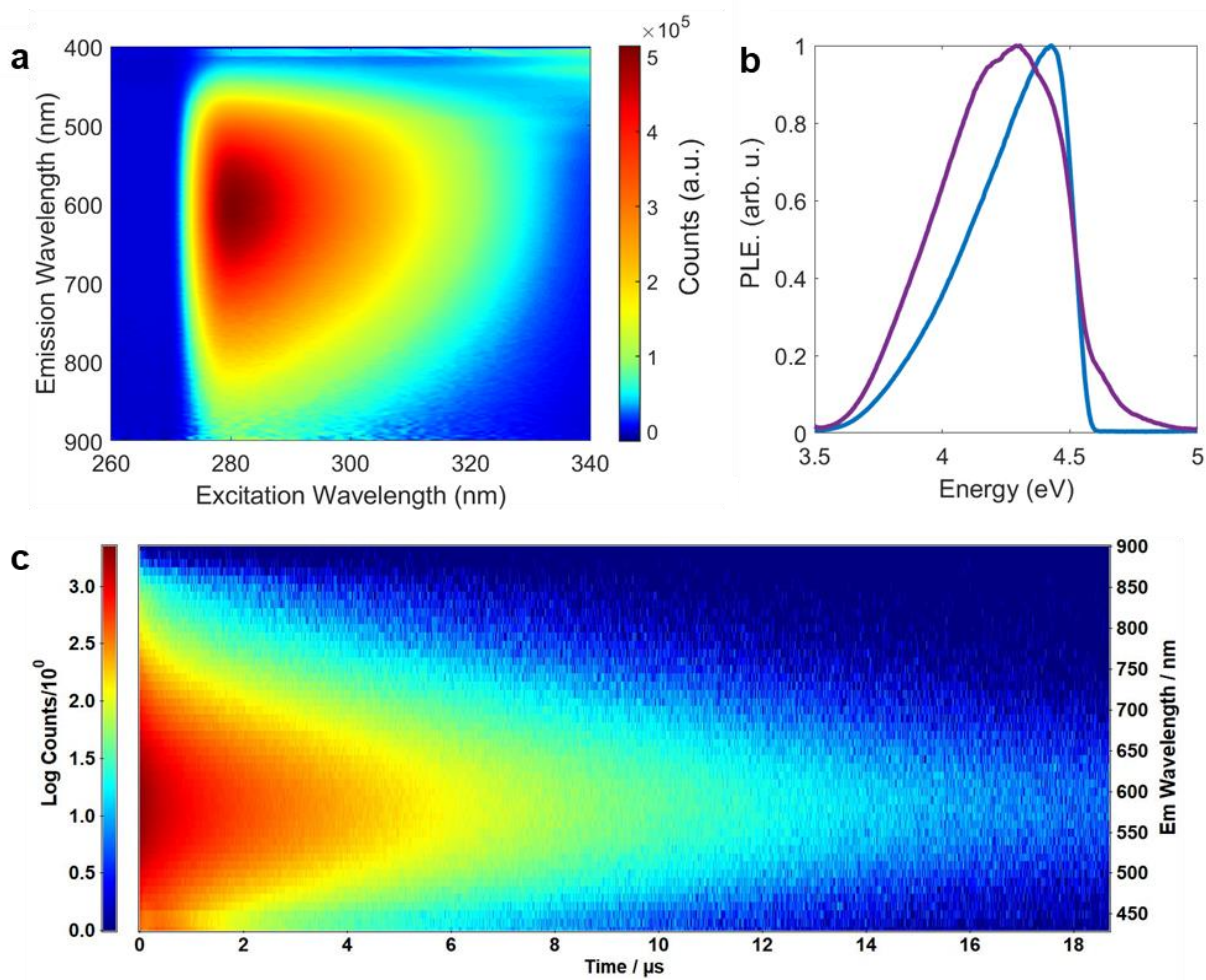

Figure S8: (a) Emission map of $Cs_2AgInCl_6$ nanoplates. (b) Photoluminescence – Excitation (PLE) spectra of $Cs_2AgInCl_6$ nanoplates (blue lines) in comparison to $Cs_2AgInCl_6$ nanocubes (purple line). The fixed emission wavelength for PL measurements for nanoplates was 600 nm, and for nanocubes 605 nm. (c) Time-resolved emission spectra (TRES) for $Cs_2AgInCl_6$ nanoplates, measured with EPL 375nm laser, showing $\tau=3.2\cdot10^{-6}$ sec with a standard deviation of $1.9\cdot10^{-8}$ sec. It is fit the long life-time typical for this type of perovskites.

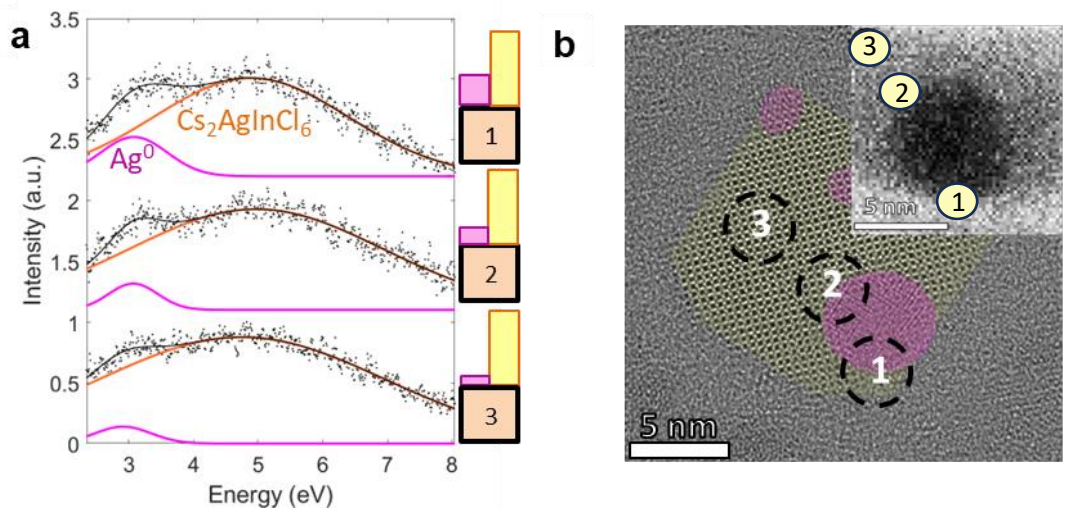

Figure S9: (a) EELS measurement from different areas on the $Cs_2AgInCl_6$ nanoplate structure illustrated by numbers in (b) the TEM micrograph. The measurement shows the $Cs_2AgInCl_6$ bandgap and the metallic silver surface plasmon peak. The purple and yellow bars show the silver surface plasmon peak amplitude and $Cs_2AgInCl_6$ bandgap peak, respectively.
The exact areas after the scan are marked in (b) inset, showing an ADF-STEM micrograph of $Cs_2AgInCl_6$ nanoplate, after the measurement. The high-contrast particle is a silver nanoparticle, located on $Cs_2AgInCl_6$ nanoplates. The different regions marked with a number indicate EELS measurements taken from (1) a silver particle on a perovskite matrix, (2) a region containing a silver particle and perovskite, and (3) a perovskite region that is a few nanometers away from silver particles.

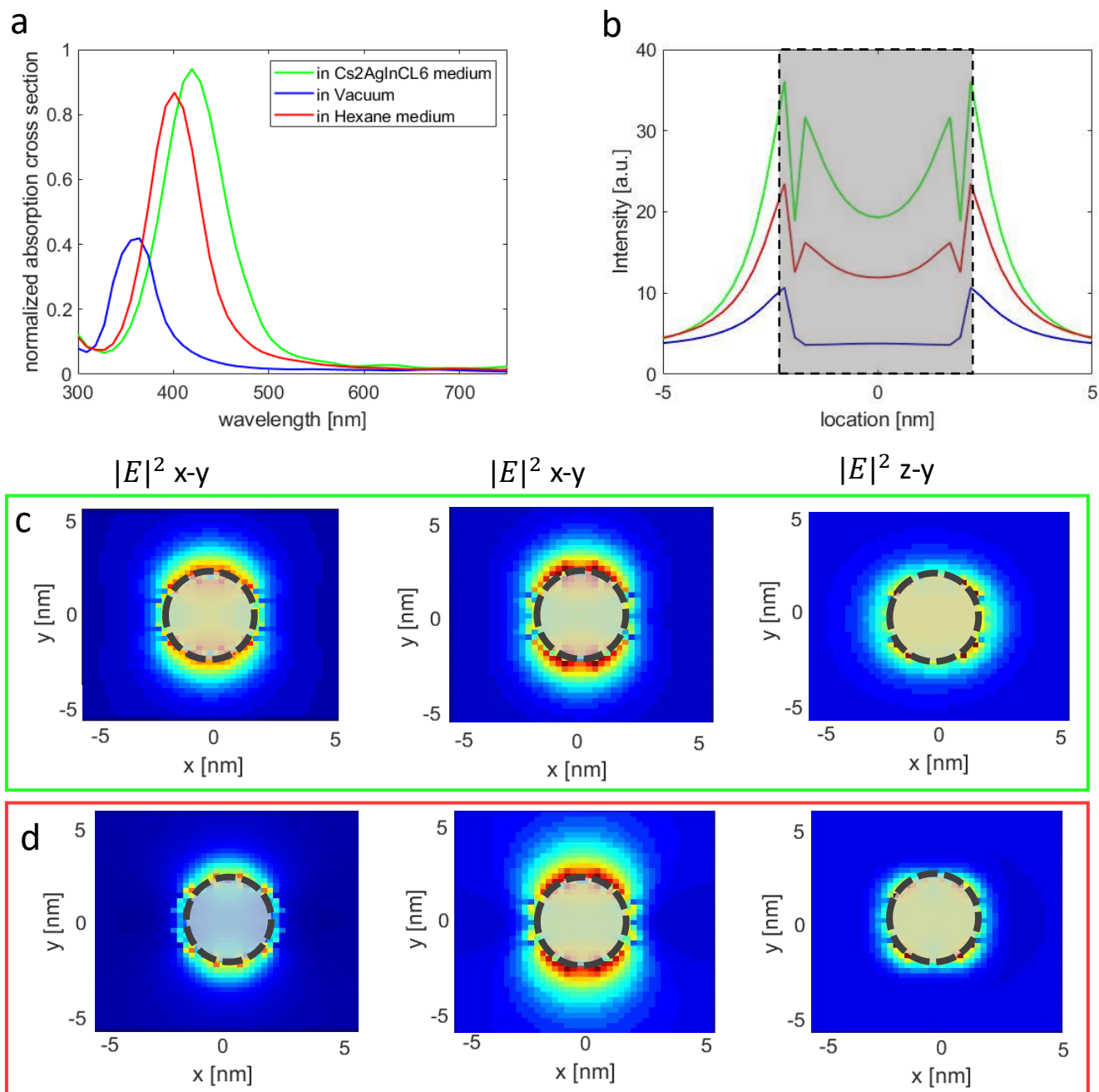

Figure S10: Simulation of the electric fields around Ag nanoparticle with 2 nm radius in different mediums. (a) absorption cross-section normalized to the particle size of Ag nanoparticle surrounded by $Cs_2AgInCl_6$ (green), Hexane (red), and in Vacuum (blue). (b) Electric field intensity profile around the Ag nanoparticle surrounded by $Cs_2AgInCl_6$ (green), Hexane (red), and in Vacuum (blue) taken at the plasmonic wavelength found in (a), the grey area represents the Ag nanoparticle location. (c) and (d) Electric field intensity profile at different planes around the nanoparticle, shown in a grey circle, in $Cs_2AgInCl_6$ surrounding in (c) and in Hexane in (d).

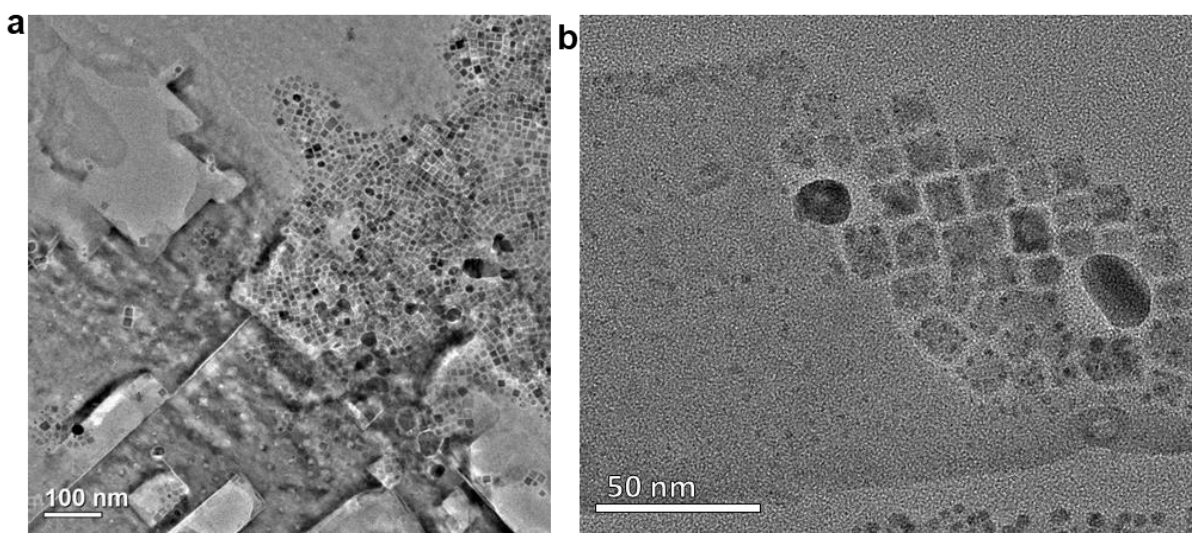

Figure S11: (a) and (b) show TEM micrograph that was taken from a typical area of $Cs_2AgInCl_6$ nanoplates sample, after a few days of storage at room temperature, showing the nanoplates fussing process to a bigger nanoplate with straight edges, on the TEM grid.

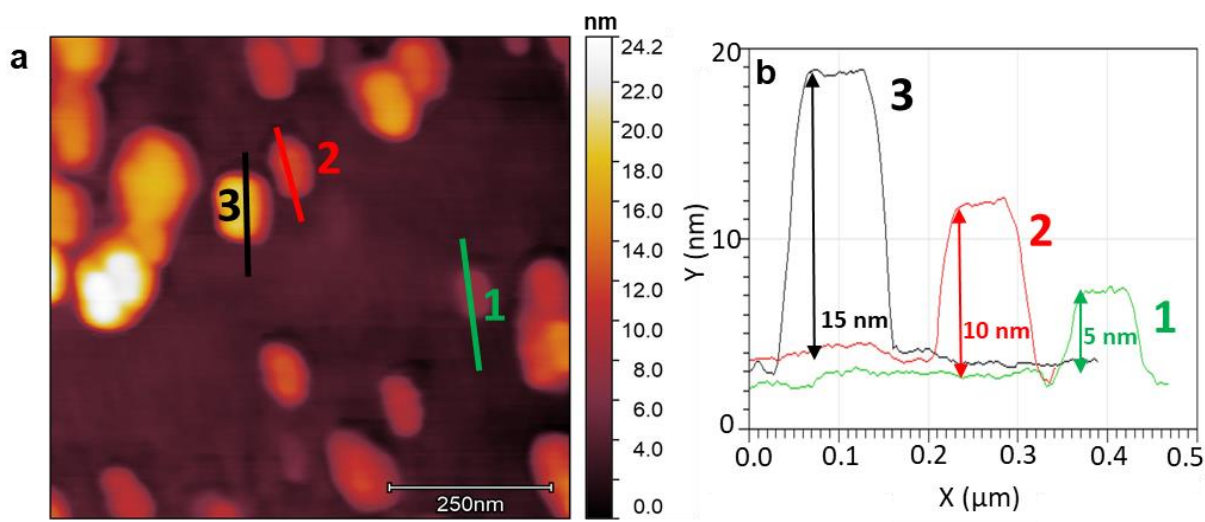

Figure S12: (a) AFM image of $Cs_2AgInCl_6$ nanoplates and (b) their height profiles showcasing uniform ~5 nanometers thickness (with ligands).

This AFM measurements confirm $Cs_2AgInCl_6$ nanoplates' thickness (including organic ligands) to be ~5nm on average, agreeing with our measurements conducted on TEM micrographs. The height profiles shown in this Figure depict the heights of a group of nanoplates with the thickness of a single plate (~5 nm), a group of nanoplates with the thickness of two plates placed on top of each other (~10 nm) and a group of nanoplates with 3 plates placed on top of each other (~15 nm). It can be seen that consistently, the thickness of one nanoplate, including organic ligands, is ~5 nm.

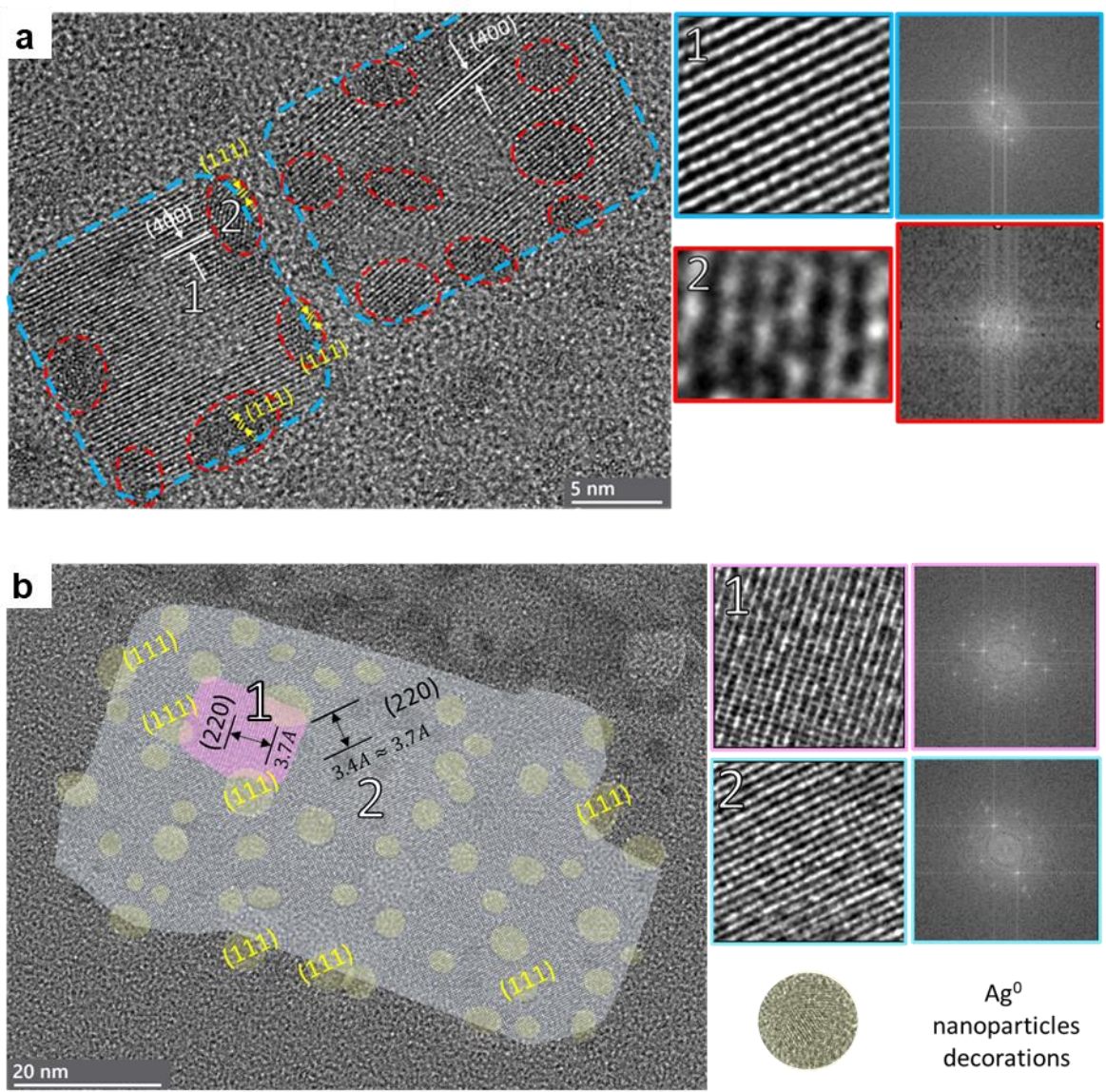

Figure S13: (a) TEM micrograph of two $Cs_2AgInCl_6$ nanoplate structure with FFT analysis from the nanoplates and the silver nanoparticle areas, showing the preferred orientation with as close as possible d-spaces for the both structures (almost epitaxial) of (400) with d-space of 2.6 Å for the $Cs_2AgInCl_6$ perovskite nanoplate and (111) d-splace of 2.4 Å for the silver nanoparticles. (b) TEM micrograph was taken from a typical area of $Cs_2AgInCl_6$ nanoplates sample, after a few days of storage at room temperature, showing the nanoplates fussing to a bigger rectangular nanoplate on the TEM grid. Inside the big nanoplate there is clearly one small nanoplate that not fussed in because it rotated relative to the large plate. The plates have similar (220) orientation with d-space of 3.7 Å and the silver nanoparticles have (111) orientation with d-space of 2.4 Å.

**Tables:**

Table S1 – PLQY of nanocubes from different syntheses.

| 1 | 2 | 3 | 4 | 5 | 6 | 7 | Average |
|---|---|---|---|---|---|---|---|
| 0.23% | 0.12% | 0.24% | 0.32% | 0.25% | 0.28% | 0.22% | 0.24% ± 0.00(1) |

Table S2 – PLQY of nanoplates from different syntheses.

| 1 | 2 | 3 | 4 | 5 | 6 | 7 | Average |
|---|---|---|---|---|---|---|---|
| 1.37% | 1.24% | 0.90% | 0.90% | 0.81% | 0.69% | 0.89% | 0.97% ± 0.00(2) |

**References**

(1) Soni, A.; Bhamu, K. C.; Sahariya, J. Investigating Effect of Strain on Electronic and Optical Properties of Lead Free Double Perovskite $Cs_2AgInCl_6$ Solar Cell Compound: A First Principle Calculation. *J Alloys Compd* **2020**, *817*, 152758.

(2) Riedle, E.; Kozma, I. Z.; Krok, P. Direct Measurement of the Group-Velocity Mismatch and Derivation of the Refractive-Index Dispersion for a Variety of Solvents in the Ultraviolet. *JOSA B, Vol. 22, Issue 7, pp. 1479-1485* **2005**, *22* (7), 1479–1485.

(3) Soma, D.; Igarashi, K.; Wakayama, Y.; Takeshima, K.; Kawaguchi, Y. *Handbook of Optical Constants of Solids I - III by E. Palik*; 2015.

(4) Levy, S.; Khalfin, S.; Pavlopoulos, N. G.; Kauffmann, Y.; Atiya, G.; Shaek, S.; Dror, S.; Shechter, R.; Bekenstein, Y. The Role Silver Nanoparticles Plays in Silver-Based Double-Perovskite Nanocrystals. *Chemistry of Materials* **2021**, *33, 7*, 2370–2377.

(5) *Liquids - Dielectric Constants*. Engineering ToolBox, (2008). Liquids - Dielectric Constants. [online] Available at: https://www.engineeringtoolbox.com/liquid-dielectric-constants-d_1263.html

(6) Pokroy, B.; Fitch, A. N.; Zolotoyabko, E. The Microstructure of Biogenic Calcite: A View by High-Resolution Synchrotron Powder Diffraction. *Advanced Materials* **2006**, *18* (18), 2363–2368.